\documentclass[10pt]{article}  % specifies document class (article) and point size (10pt)
\pdfoutput=1
\usepackage[utf8x]{inputenc}
\usepackage{bm}
\usepackage {amsthm,amsmath,amssymb,amsfonts,amssymb}
\usepackage{listings} 
\usepackage{booktabs}
\usepackage{array}
\usepackage{enumerate}
\usepackage{graphicx}  
\usepackage{caption}
\usepackage{subcaption}
\usepackage{enumitem}
\usepackage{authblk}
\usepackage{lmodern}
\usepackage[left=1in,right=1in,top=1.5in,bottom=1.5in]{geometry}
\usepackage{setspace} 
\usepackage{natbib}
\usepackage[draft]{fixme}
\usepackage{mathtools}
\usepackage{hyphenat}
\usepackage[section]{placeins}
\usepackage{xcolor}
\usepackage{bbm} % bold digits in math mode via \mathbbm

\DeclareMathOperator{\Var}{Var}         % Variance
\DeclareMathOperator{\E}{\mathbb{E}}    % Expected value
\usepackage{url}



\usepackage{dcolumn}
\newcolumntype{e}{D{.}{.}{9}}

\numberwithin{equation}{section}

\theoremstyle{plain}

\theoremstyle{definition}

\theoremstyle{remark}

\title{Modelling corporate defaults: A Markov-switching Poisson log-linear autoregressive model}  % specifies big, fancy title
\author[1]{Geir D. Berentsen}
\author[1, 2]{Jan Bulla}
\author[3]{Antonello Maruotti}
\author[1]{B\r{a}rd St\o ve}

\affil[1]{Department of Mathematics, University of Bergen, P.O. Box 7803, 5020 Bergen, Norway}
\affil[2]{Department of Psychiatry and Psychotherapy, University Regensburg, Universit\"atsstra\ss{}e 84, 93053 Regensburg, Germany}
\affil[3]{Dipartimento di Giurisprudenza, Economia, Politica e Lingue Moderne (GEPLI), Libera Universit\`a Maria Ss Assunta, Via Pompeo Magno 22, 00192 Rome, Italy}

\begin{document} 
%%%%%%%%%%%%%%%%%%%%%%%%%%%%%%%%%%%%%%%%%%%%%%%%%%%%%%%%%%%%%%%%%%%%%%%

\maketitle

\abstract{This article extends the autoregressive count time series model class by allowing for a model with regimes, that is, some of the parameters in the model depend on the state of an unobserved Markov chain. We develop a quasi-maximum likelihood estimator by adapting the extended Hamilton-Grey algorithm for the Poisson log-linear autoregressive model, and we perform a simulation study to check the finite sample behaviour of the estimator. The motivation for the model comes from the study of corporate defaults, in particular the study of default clustering. We provide evidence that time series of counts of US monthly corporate defaults consists of two regimes and that the so-called contagion effect, that is current defaults affect the probability of other firms defaulting in the future, is present in one of these regimes, even after controlling for financial and economic covariates. We further find evidence for that the covariate effects are different in each of the two regimes. Our results imply that the notion of contagion in the default count process is time-dependent, and thus more dynamic than previously believed.}\\


%%%%%%%%%%%%%%%%%%%%%%%%%%%%%%%%%%%%%%%%%%%%%%%%%%%%%%%%%%%%%%%%%
\section{Introduction}
\label{sec:intro}
%%%%%%%%%%%%%%%%%%%%%%%%%%%%%%%%%%%%%%%%%%%%%%%%%%%%%%%%%%%%%%%%%

The study of modeling and forecasting corporate defaults has been intensified in the recent years. A major drive for this increased interest has been the need to find an explanation for the clustering of defaults observed. In essence, two explanations have been proposed for this stylized fact. First, each firm can be considered exposed to a "systematic risk", represented by common economical and financial factors. Second, one firm's default may increase the likelihood of other firms defaulting, resulting in so-called "contagion effects". Both explanations are plausible approaches describing a clustering of the observed defaults, and may occur separately but also jointly.
    
From a practical perspective, one core issue has been to distinguish between these two explanations. In particular, revelation of the presence of a contagion effect is important: disregarding this effect may lead to an underestimation of probabilities of defaults (PD), because most credit models in practice assume that default events are conditionally independent (i.e.~given observable common factors, defaults are independent in time). Consequently, ignoring possible dependence structures in corporate defaults may cause that the amount of capital held by banks and other financial institutions exposed to credit portfolios is insufficient.

Several studies have examined the default clustering fact in the past, and following \cite{agosto2016}, one can broadly divide the studies into two categories.\\
In the first category, firm-level data are available in addition to macroeconomic variables and default times for firms are recorded. Then, the default times are usually modeled by Poisson processes with both types of covariates entering the default intensities. Studies in this category are e.g.~\cite{das2007}, who provide evidence that the "systemic risk" on its own cannot explain the degree of clustering observed in U.S.~industrial defaults, and \cite{lando2010}, who does not report a contagion effect using another type of test procedure.\\
The second category uses aggregate data, where the number of defaults in a given time period is collected together with macroeconomic variables. Several papers have used this approach, e.g.~\cite{koopman2012}, who use a high-dimensional and partly
nonlinear, non-Gaussian dynamic factor model for counts of default, where the probability of default is time-varying and a function of macroeconomic covariates. However, their model specification requires computationally demanding Monte Carlo methods. They find that the extreme tail clustering in defaults cannot be captured using macro variables alone. Another study in the second category was performed by \cite{azizpour2017}. They find strong evidence that contagion is a main source for the clustering behavior.\\
The papers most related to our approach are \cite{agosto2016} and \cite{santanna2017}, both of which belong to the second category as well. \cite{agosto2016} introduce a class of Poisson autoregressive models with exogenous covariates to model the count of defaults. They find evidence of a contagion effect, which diminishes in recent years. \cite{santanna2017} introduces new test procedures permitting to carry out model checks for dynamic count models, and finds evidence of a contagion effect as well.

In this paper we propose an extension to the count time series model class. We allow for a model that can characterize the time series behaviors in different regimes. By permitting switching between these regime, such a model is able to capture more complex dynamic patterns. In essence, we have a model where (some of) the parameters depend on the state of an unobserved Markov chain. There has been some work in this direction, confer e.g. \cite{macdonald2016}. This also means that the macroeconomic and financial variables can have different effects on the default intensity depending on the regime. This extension is inspired by some of the results in \cite{agosto2016}, in where they show that there are structural instabilities in the model parameters over the sample period. We should be able to pick up such effects by permitting for several regimes. We further will be able to provide more evidence for or against the contagion effect. 

In the past years models for count time series have been intensively studied, in particular due to the wide area of applications. This paper contributes to the literature on autoregressive models for count time series subject to structural changes. Time series often experiences a structural change, and the problem of change point detection has been a central issue in the literature: see e.g.~\cite{kirch2016} for a recent review. The change point test for univariate integer-valued time series has been studied by many authors, see e.g. \cite{kang2009}, \cite{franke2012}, \cite{fokianos2014}, \cite{kang2014}, \cite{doukhan2015}  and \cite{diop2017}, while a procedure on testing for bivariate models is given in \cite{lee2016}. A natural extension will be to allow for a count time series model with regimes, and to the best of our knowledge, this is the first paper employing such an idea in the setting of autoregressive count time series.    

 The paper is organized as follows. In Section 2, we extend the Poisson log-linear autoregressive count time series model allowing for regime-switching, and interpret it in the context of modeling defaults. In Section 3 we present the algorithm for how the model can be estimated, and adress both inference about the underlying regimes and prediction. In Section 4 we provide a simulation study for assessing the performance of the algorithm, and perform  the empirical analysis of the counts of corporate defaults. Finally, Section 5 provides some concluding remarks. 

%{\color{red}
%We probably need to write more in details regarding our definition of contagion etc - as section 2.1 and 2.2 in \cite{agosto2016}. Perhaps in section 2? Or with the application?} 

%%%%%%%%%%%%%%%%%%%%%%%%%%%%%%%%%%%%%%%%%%%%%%%%%%%%%%%%%%%%%%%%%%%%%%%%%%%%%%%%%%%%%%%%%%
\section{The Markov-switching Poisson log-linear autoregressive model}
%%%%%%%%%%%%%%%%%%%%%%%%%%%%%%%%%%%%%%%%%%%%%%%%%%%%%%%%%%%%%%%%%%%%%%%%%%%%%%%%%%%%%%%%%%

In this section we define the Markov-switching Poisson log-linear autoregressive (MS-PLLAR) model and interpret it in the context of modeling defaults.

%{\color{red}: Introduce the abbreviation MS-PLLAR here?}
%%%%%%%%%%%%%%%%%%%%%%%%%%%%%%%%%%%%%%%%%%%%%%%%%%%%%%%%%%%%%%%%%%%%%%%%%%%%%%%%%%%%%%%%%%
\subsection{Definition}
%%%%%%%%%%%%%%%%%%%%%%%%%%%%%%%%%%%%%%%%%%%%%%%%%%%%%%%%%%%%%%%%%%%%%%%%%%%%%%%%%%%%%%%%%%

Let $\{Y_t\}$ be a count time series such as corporate defaults, and let  $\{X_t\}$ denote a r-dimensional time-varying exogenous covariate vector, i.e.~${X}_t = (X_{t,1},\dots, X_{t,r})^t$. To capture possible regime changes in $\{Y_t\}$ we introduce an unobserved first-order Markov process $\{S_t\}$ taking discrete values $1, \dots, m$. Let $\Gamma = \{\gamma_{ij}\}$ denote the $m\times m$ transition probability matrix of $\{S_t\}$, where the terms $\{\gamma_{ij}\}$ represent the probability of moving from the $\mbox{i}^{ \mbox{\tiny th}}$ state at time $t-1$ to state $j$ at time $t$, where $h, k = 1,\dots, m$. We assume that $S_t$ is time-homogeneous and stationary, with $\delta = (\delta_1,\dots,\delta_m)$ denoting the stationary distribution. Furthermore, $Y^{(t)}$ represents the vector of observations $(Y_1, \dots, Y_t)^t$, and the vector of hidden states $S^{(t)}$ is defined analogously. Similarly, $X^{(t)}$ denotes the matrix of covariates $(X_1, \dots, X_t)^t$ and $\beta_{S_t}$ denotes the vector of corresponding state-specific effects $(\beta_{1, S_t},\dots, \beta_{r, S_t})^t$. For modeling corporate defaults we consider a MS extension of the Poisson log-linear autoregressive model studied in \cite{fokianos2011} defined by
\begin{equation}
\label{eq:msloglinear}
Y_t\mid \mathcal{F}_{t - 1}\sim \text{Poisson}(\lambda_{t}),\quad\eta_t = \log(\lambda_{t}) = d_{S_t} + a_{S_t} \eta_{t-1} + b_{S_t} \log(Y_{t-1} + 1) + \beta_{S_t}^\prime X_t, \quad t\geq 1
\end{equation}

with information set $\mathcal{F}_{t} = \{Y^{(t)},X^{(t + 1)}, S^{(t + 1)}, \theta\}$ and $\theta$ denoting the vector of parameters in the model, i.e.~$\theta = \left(\{a_k, b_k, d_k, \beta_{1,k},\dots,\beta_{r,k}\}_{k = 1}^m, \{\gamma_{ij}\}_{i,j = 1}^m\right)$. Since $\sum_{j = 1}^m \gamma_{ij} = 1$, the model contains $3m + rm + m(m-1)$ free parameters.   

% The idea is to let $\{S_t\}$ govern how $\lambda_t$ (and thus $\{Y_t\}$) should depend on the exogenous covariates $\{X_t\}$ and past values of $\lambda_t$ and the observed process $\{Y_t\}$. Specifically,

%%%%%%%%%%%%%%%%%%%%%%%%%%%%%%%%%%%%%%%%%%%%%%%%%%%%%%%%%%%%%%%%%%%%%%%%%%%%%%%%%%%%%%%%%%
\subsection{Interpretation of the model}\label{sec:interpretation}
%%%%%%%%%%%%%%%%%%%%%%%%%%%%%%%%%%%%%%%%%%%%%%%%%%%%%%%%%%%%%%%%%%%%%%%%%%%%%%%%%%%%%%%%%%

For $m = 1$ \eqref{eq:msloglinear} reduces to the log-linear autoregressive model of \cite{fokianos2011}. That is, the linear predictor $\eta_t$ reduces to 
\begin{equation}
\label{eq:loglinear_simple}
\eta_t = \log(\lambda_{t}) = d + a \eta_{t-1} + b \log(Y_{t-1} + 1) + \beta^\prime X_t.
\end{equation}
Note that in this framework, we can allow for $\beta^\prime X_t < 0$. 
The roles of the terms of \eqref{eq:loglinear_simple} can be interpreted as follows. First, the parameter $d$ simply fixes the overall intensity level. Then, changes in the systematic risk are best captured by the term $\beta^\prime X_t$, which models the impact of exogenous variables representing macroeconomic or financial risks on the (log-)intensity. Moreover, assuming $b > 0$, the second-last term  of \eqref{eq:loglinear_simple}, $b \log(Y_{t-1})$,  may indicate the presence of contagion effects since increases in $Y_{t-1}$ lead to an intensity increase. The term $a \eta_{t-1}$ is slightly less straightforward to interpret. On the one hand, it may either have an amplifying effect on high intensity values (for $a > 0$), or dampen extreme values (for $a < 0$). On the other hand, $a \eta_{t-1}$ also implicitly models a dependence of the intensity on all previous lags of both $Y_{t}$ and exogenous variables $X_t$. This is best illustrated by assuming for simplicity that $Y_0$ and $\eta_0$ are known quantities. Assuming $a \neq 1$, we then obtain
\begin{equation}
\label{eq:loglinear_advanced}
\eta_t = d\frac{1 - a^t}{1-a} + a^t\eta_0 + b\sum_{i=0}^{t-1}a^i\log(1+Y_{t-i-1}) + \sum_{i=0}^{t-1}a^i\beta^{\prime}X_{t-i}
\end{equation}
by repeated substitution of \eqref{eq:loglinear_simple}. From Equation \eqref{eq:loglinear_advanced} we see that all terms related to systematic risk propagate to future values of the (log-)intensity as $\sum_{i=0}^{t-1}a^i\beta^{\prime}X_{t-i}$, which we can therefore interpret as a term representing the overall macroeconomic or financial risk. Similarly, the terms linked to contagion effects sum up to $b\sum_{i=0}^{t-1}a^i\log(1+Y_{t-i-1})$, and thus propagate to future values of the (log-)intensity as overall feedback-effect, i.e.~the overall contagion. Summarizing, the inclusion of the term $a \eta_{t-1}$ in \eqref{eq:loglinear_simple} represents a parsimonious way for allowing the intensity to depend on all previous lags of both $Y_{t}$ and exogenous variables $X_t$.\\
For the MS case, i.e.~$m \geq 2$, the interpretation of the last three terms of \eqref{eq:msloglinear} are similar to the simple case. However, the model coefficients are driven by the unobserved Markov chain, which permits more flexibility since it allows the above described effects to vary in time. For example, exogenous variables may have a significant impact on the intensity in one state, while these effects remain negligible in another state. Moreover, the first parameter of \eqref{eq:loglinear_simple} becomes $d_{S_t}$ and permits to model changes in the systematic risk, resulting e.g.~from unobserved covariates.

To summarize; we thus follow \cite{agosto2016} for allowing for differentiation between systematic risk and contagion, and we note that in the case where $b = 0$ in equation \eqref{eq:loglinear_simple}, the model imply conditional independence between current and past defaults. Similarly, in the MS case, we will examine the estimated parameters of $b_{S_t}$ for all regimes. Thus, we may actually observe that in some regimes we may have contagion, and in others not. We thus allow for a more dynamic process of the corporate default counts, but retain an analogous interpretability as the PARX model of \cite{agosto2016}.

%Thus, the inclusion of the term $a \eta_{t-1}$ is the most parsimonious way one can let the intensity depend on all previous lags of both $Y_{t}$ and exogenous variables $X_t$. Changes in the systematic risk are best captured by the term $\beta^\prime X_t$, which models the impact of exogenous variables representing macroeconomic or financial risks on the (log-)intensity. 
%Assuming $b > 0$, the second-last term  of \eqref{eq:loglinear_simple}, $b \log(Y_{t-1})$, may indicate the presence of contagion effects since increases in $Y_{t-1}$ lead to an intensity increase. Similarly, these terms propagate to future values of the (log-)intensity as $b\sum_{i=0}^{t-1}a^i\log(1+Y_{t-i-1})$, which represent the overall feedback-effect, i.e. the overall contagion. Last, the parameter $d$ simply fixes the overall intensity level.\\       

%%%%%%%%%%%%%%%%%%%%%%%%%%%%%%%%%%%%%%%%%%%%%%%%%%%%%%%%%%%%%%%%%%%%%%%%%%%%%%%%%%%%%%%%%%
\section{Estimation and inference}
\label{sec:estimation}
%%%%%%%%%%%%%%%%%%%%%%%%%%%%%%%%%%%%%%%%%%%%%%%%%%%%%%%%%%%%%%%%%%%%%%%%%%%%%%%%%%%%%%%%%%

In this section we present the algorithm for estimating the model parameters, address inference about the underlying regimes, and derive a couple of prediction techniques.

%%%%%%%%%%%%%%%%%%%%%%%%%%%%%%%%%%%%%%%%%%%%%%%%%%%%%%%%%%%%%%%%%%%%%%%%%%%%%%%%%%%%%%%%%%
\subsection{The regime path dependence problem}
%%%%%%%%%%%%%%%%%%%%%%%%%%%%%%%%%%%%%%%%%%%%%%%%%%%%%%%%%%%%%%%%%%%%%%%%%%%%%%%%%%%%%%%%%%

The computation of $\eta_t$ in \eqref{eq:msloglinear} requires the comprehensive information set $\mathcal{F}_{t - 1} = \{Y^{(t - 1)},X^{(t)}, S^{(t)}, \theta\}$  due to its dependence on past values of $\eta_t$. In particular, $\eta_t$ depends on the complete regime path $S^{(t)}$, which causes difficulties within the estimation procedure. The likelihood, denoted by $L(\theta)$ of the observations $\{Y^{(T)}\}$, is given by
\begin{equation}\label{eq:lik1}
\begin{split}
L(\theta) &= P(Y^{(T)} = y^{(T)}\mid \theta) = \sum_{s_1, \dots, s_T = 1}^m P(Y^{(T)} = y^{(T)}\mid S^{(T)} = s^{(T)})P(S^{(T)} = s^{(T)})\\
& = \sum_{s_1, \dots, s_T = 1}^m \left(\prod_{t = 1}^T \frac{\exp(-\lambda_{t})\lambda_{t}^{y_t}}{y_{t}!}P(S_1)\prod_{t = 2}^T P(S_t = s_t\mid  S_{t-1} = s_{t-1})\right).
\end{split}
\end{equation}
A direct computation of \eqref{eq:lik1} is problematic since $\lambda_t = \exp(\eta_t)$ has to be derived recursively by \eqref{eq:msloglinear} for each of the $m^T$ different regime paths. As a consequence, direct computation of \eqref{eq:lik1} quickly becomes infeasible with increasing $T$. This problem is often termed the path-dependence problems for Markov-switching (MS) models, and was first pointed out by \cite{hamilton1994} when discussing the possibility of a MS generalized autoregressive conditional heteroskedasticity (MS-GARCH) model. The problem for the MS-GARCH model was later adressed in the work of \cite{gray1996}, see also \cite{augustyniak2014} and references therein.\\

In the seminal work of \cite{hamilton1989} a much used algorithm for estimating MS autoregressive (MS-AR) models is proposed. However, for MS autoregressive moving-average (MS-ARMA) models path-dependence problems arise due to the moving-average component \citep[see e.g.][]{billio1998}, in which case the algorithm of Hamilton fails. Analogously, with the exception of the case $a = 0$ in \eqref{eq:msloglinear}, a direct adaptation of the Hamilton algorithm to the MS-PLLAR model faces path-dependence type problems. Such difficulties are the very ones addressed by the principles presented in \cite{gray1996}, which build the foundation for our approach. More precisely, in this paper we approximate \eqref{eq:lik1} using an adaptation of the extended Hamilton-Gray (EHG) algorithm described in \cite{chen2011}. The EHG algorithm avoids the path dependence problem by combining the algorithm of \cite{hamilton1989} and the ideas of \cite{gray1996} of recursively replacing certain quantities, in our case $\eta_{t}$, with its expectation. This permits to trace only the $m^2$ possible regime paths from time $t - 1$ to time $t$ instead of the full path, and then to iteratively replace $\eta_t$ with the corresponding conditional expectations that are consistent with these paths. Hence, the estimation routine falls into the framework of \cite{hamilton1989}. In order to separate our adaption from the original EHG algorithm tailored for MS-ARMA models, we refer to the adaptation as the MS-PLLAR EHG (or only EHG in short) algorithm.

% At time $t$ the information available is $\Omega_{t-1}$ for which we want to calculate $P(Y_t = y_t\mid \Omega_{t-1})$ and then updated the information set to $\Omega_t$.
% The algorithm aims to filter out the sequence $\{\eta_t}$, $t = 1, \dots, T$. 
% In the model of \cite{fokianos2011}, $Y_t\mid \mathcal{F}^*_{t - 1}\sim \text{Poisson}(\lambda_{t})$, with a log-linear specification of $\lambda_t$:
% \begin{equation}
% \label{eq:loglinear}
% \eta_t = \log(\lambda_{t}) = d + a \eta_{t-1} + b \log(Y_{t-1} + 1) + \beta^t X_t , \quad t\geq 1
% \end{equation}

%%%%%%%%%%%%%%%%%%%%%%%%%%%%%%%%%%%%%%%%%%%%%%%%%%%%%%%%%%%%%%%%%%%%%%%%%%%%%%%%%%%%%%%%%%
\subsection{The MS-PLLAR EHG algorithm}
\label{sec:ehg_algo}
%%%%%%%%%%%%%%%%%%%%%%%%%%%%%%%%%%%%%%%%%%%%%%%%%%%%%%%%%%%%%%%%%%%%%%%%%%%%%%%%%%%%%%%%%%

% For later: Maybe mention somewhere in this section already that the algorithm provides estimates via QMLE?

\begin{figure}[ht]
        \centering
\includegraphics[scale=0.75]{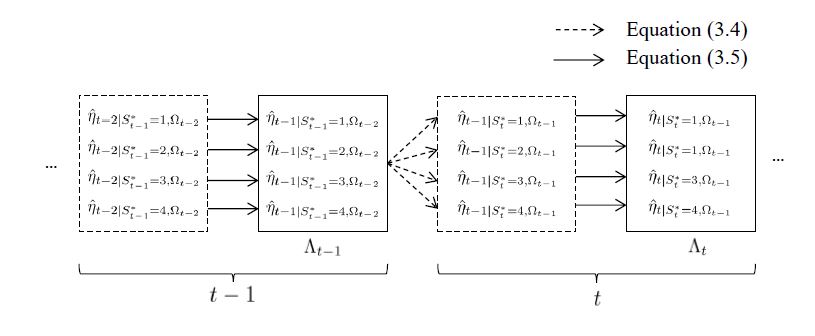}
       \caption{The figure shows the evolution of $\Lambda_t$ when there are $m = 2$ states. Note that once $\Lambda_t$ is obtained, the state filtering defined by equations \eqref{eq: ycond} - \eqref{eq:onestepahead} must be employed before proceeding to t + 1. }
        \label{fig:ehgflow}
\end{figure}

For tracing the regime path from time $t - 1$ to time $t$, we create a new state variable $S^{*}_t = 1, 2, \dots, m^2$. This variable is defined such that each state of $S^{*}_t$ represent a particular regime path $(S_{t-1} = s_{t-1}, S_{t} = s_{t})$, i.e.
\begin{equation}\label{Sstar}
S^{*}_t = \left\{
	\begin{array}{ccc}
1      &   \text{if} & \quad S_t = 1,\quad S_{t-1} = 1 \\
2      &   \text{if} & \quad S_t = 2,\quad S_{t-1} = 1 \\
\vdots &             & \vdots\\
m^2    &   \text{if} & \quad S_t = m,\quad S_{t-1} = m 
	\end{array}
\right.
\end{equation}
Note that $S^{*}_t$ also inherits the Markov property from $S_t$. The dynamics of $S^{*}_t$ can be characterized via a first-order Markov chain with a $m^2\times m^2$ transition probability matrix $\Gamma^* = \{\gamma^*_{kh}\}$ that can be derived from $\Gamma$. For example, for $m = 2$ one obtains
\begin{equation*}
\Gamma^* =
\begin{bmatrix} 
 \gamma^*_{11} & \gamma^*_{12} & \gamma^*_{13} & \gamma^*_{14} \\
 \gamma^*_{21} & \gamma^*_{22} & \gamma^*_{23} & \gamma^*_{24} \\
 \gamma^*_{31} & \gamma^*_{32} & \gamma^*_{33} & \gamma^*_{34} \\
 \gamma^*_{41} & \gamma^*_{42} & \gamma^*_{43} & \gamma^*_{44} 
 \end{bmatrix}
 = 
 \begin{bmatrix} 
 \gamma_{11} & \gamma_{12} & 0 & 0 \\
 0 & 0 & \gamma_{21} & \gamma_{22} \\
 \gamma_{11} & \gamma_{12} & 0 & 0 \\
 0 & 0 & \gamma_{21} & \gamma_{22} 
 \end{bmatrix}.
\end{equation*}
The new state variable $S^{*}_t$ is crucial for determining the aforementioned conditional expectations of $\eta_t$. For each time $t$, the conditional expectations will be collected in a vector denoted by $\Lambda_t$. The computation of $\Lambda_t$  bases on a specific information set denoted by $\Omega_{t-1}$ as well as $\Lambda_{t-1}$, which are given by
\begin{eqnarray}
\label{eq:iset}
\Omega_{t-1} &=& \{Y^{(t-1)}, X^{(t)}, \Lambda_{t-1}, \theta\}, \quad \nonumber \\ 
\Lambda_{t-1} &=& \left(\hat{\eta}_{t - 1\mid S^*_{t - 1} = 1, \Omega_{t-2}}, \hat{\eta}_{t - 1\mid S^*_{t - 1} = 2, \Omega_{t-2}},
\dots, \hat{\eta}_{t - 1\mid S^*_{t - 1} = m^2, \Omega_{t-2}}\right).
\end{eqnarray}
Thus, the vector $\Lambda_{t - 1}$ contains the corresponding expectations of $\eta_{t-1}$ conditional on $S^*_{t - 1} = j$, $j = 1, \dots, m^2$ and the information set $\Omega_{t-2}$, indicating the recursive structure of the algorithm. The first step in deriving $\Lambda_t$ is to compute the expectations of the elements in $\Lambda_{t-1}$ conditional on $S^*_{t} = j$, $j = 1, \dots, m^2$ and the information set $\Omega_{t-1}$ by
% so that  $\hat{\eta}_{t-1\mid S^*_{t} = j, \Omega_{t-2}}$ is updated with the information $\Omega_{t-1}$ to $\hat{\eta}_{t-1\mid S^*_{t} = j, \Omega_{t-1}}$:
\begin{equation}\label{eq:lambda1}
\begin{split}
\hat{\eta}_{t-1\mid S^*_{t} = j, \Omega_{t-1}} &=  E(\hat{\eta}_{t-1\mid S^*_{t-1}, \Omega_{t-2}} \mid S^*_{t} = j, \Omega_{t-1})\\ 
& = \sum_{i=1}^{m^2} P(S^*_{t - 1} = i \mid S^*_{t} = j, \Omega_{t-1})\hat{\eta}_{t - 1\mid S^*_{t-1} = i, \Omega_{t-2}}\\
& = \sum_{i=1}^{m^2}\frac{ \gamma^*_{ij}P(S^*_{t-1} = i \mid \Omega_{t-1})\hat{\eta}_{t - 1\mid S^*_{t-1} = i, \Omega_{t-2}}}{P(S^*_{t} = j \mid \Omega_{t-1})},
% & = \frac{\sum_{i=1}^{m^2} P(S^*_{t} = j \mid S^*_{t-1} = i)P(S^*_{t-1} = i \mid \Omega_{t-1})\hat{\eta}_{t - 1\mid S^*_{t-1} = i , \Omega_{t-2}}}{\sum_{i=1}^{m^2} P(S^*_{t} = j \mid S^*_{t-1} = i)P(S^*_{t-1} = i \mid \Omega_{t-1})}\\
% j = 1, \dots m^2$.
\end{split}
\end{equation}
where $j = 1, \dots, m^2$. The substitution of $P(S^*_{t} = j \mid S^*_{t-1} = i, \Omega_{t-1})$ with $P(S^*_{t} = j \mid S^*_{t-1} = i) = \gamma^*_{ij}$ is valid due to the Markov property of $S^*_{t}$. This step can be seen as a Bayesian update of the elements of $\Lambda_{t-1}$ with the information set $\Omega_{t-1}$. For the next step, let $S_t(S^*_{t} = j)$ be the value of $S_t$ given that $S^*_{t}$ is in state $j$. Since each state of $S^*_{t}$ represent a particular realization of $(S_{t-1}, S_t)$, the elements of $\Lambda_t$ can be computed by forwarding $\hat{\eta}_{t-1\mid S^*_{t} = j, \Omega_{t-1}}$ consistently with the regime path corresponding to $S^*_{t} = j$, $j = 1, \dots, m^2$:
\begin{equation}
\label{eq:lambda2}
\begin{split}
\hat{\eta}_{t\mid S^*_{t} = j, \Omega_{t-1}} & = E(\eta_{t} \mid S^*_{t} = j, \Omega_{t-1}  )\\
 & = d_{S_t(S^*_{t} = j)} + a_{S_t(S^*_{t} = j)}\hat{\eta}_{t - 1\mid S^*_{t} = j, \Omega_{t-1}} + b_{S_t(S^*_{t} = j)}Y_{t-1} + \beta_{S_t(S^*_{t} = j)}^t X_t, \quad j = 1, \dots, m^2\\
\end{split}
\end{equation}
Hence, it is possible to compute $\Lambda_t$ by means of the equations \eqref{eq:lambda1}-\eqref{eq:lambda2}, provided that of the quantities $P(S^*_{t-1} = j \mid \Omega_{t-1})$  and $P(S^*_{t} = i \mid \Omega_{t-1})$, $i,j = 1, \dots, m^2$ occurring in \eqref{eq:lambda1} are known. One may note that these two quantities are already a by-product of the previous iteration step (carried out for $t-1$) through the filter defined by the equations \eqref{eq: ycond} - \eqref{eq:onestepahead} below. In detail, under the Poisson assumption the probability of $Y_t = y_t$ conditional on $S^*_{t}$ and $\Omega_{t-1}$ is
\begin{equation}
\label{eq: ycond}
P(Y_t = y_t \mid S^*_{t} = j, \Omega_{t-1}) =  \frac{\left(\hat{\lambda}_{t\mid S^*_{t} = j, \Omega_{t-1}}\right)^{y_t} \exp\left(-\hat{\lambda}_{t\mid S^*_{t} = j, \Omega_{t-1}}\right)}{y_t!},
\end{equation}
where $\hat{\lambda}_{t\mid S^*_{t} = j, \Omega_{t-1}} = \exp(\hat{\eta}_{t\mid S^*_{t} = j, \Omega_{t-1}})$. By summing over all states, the probability of $Y_t = y_t$ conditional on $\Omega_{t-1}$ then becomes
\begin{equation}\label{eq:yuncond}
P(Y_t = y_t \mid \Omega_{t-1}) = \sum_{i = 1}^{m^2} P(Y_t = y_t \mid S^*_{t} = i, \Omega_{t-1}) P(S^*_{t} = i \mid \Omega_{t-1}),   
\end{equation}
 which effectively corresponds to a discrete mixture of Poisson-distributed variables. Subsequently, the so-called filtering probabilities can be computed by 
\begin{equation}\label{eq:filtprob}
P(S^*_{t} = j\mid\Omega_{t}) = \frac{P(S^ *_{t} = j\mid\Omega_{t - 1})P(Y_t = y_t \mid S^*_{t} = j, \Omega_{t-1})}{P(Y_t = y_t \mid \Omega_{t-1})} 
\end{equation}
for $j = 1, \dots, m^2$, and the one-step ahead predictive probabilities trough
\begin{equation}
\label{eq:onestepahead}
\begin{split}
P(S^*_{t + 1} = j\mid\Omega_{t}) &= \sum_{i = 1}^{m^2} P(S^*_{t + 1} = j \mid S^*_{t} = i, \Omega_{t})P(S^*_{t} = i\mid\Omega_{t})\\
& =\sum_{i = 1}^{m^2} \gamma^*_{ij}P(S^*_{t} = i\mid\Omega_{t})
\end{split}
\end{equation}
for  $j = 1, \dots, m^2$, where the last equality is a direct consequence of the Markov property of $\{S^*_{t}\}$. Last, by recursively computing equations \eqref{eq:lambda1}-\eqref{eq:onestepahead} we can obtain the quasi-log-likelihood 
\begin{equation}
\label{eq:qll}
\log L^*(\theta) = \sum_{t = 1}^T \log P(Y_t = y_t | \Omega_{t-1}),
\end{equation}
where  $P(Y_t = y_t | \Omega_{t-1})$ is given by \eqref{eq:yuncond}. Figure \ref{fig:ehgflow} displays an overview of the evolution of $\Lambda_t$ for $m = 2$. Intuitively, the two equations \eqref{eq:filtprob} and \eqref{eq:onestepahead} serve as adaptive inference tool for $S^*_{t}$. On the one hand, the one-step ahead probability $P(S^*_{t}\mid \Omega_{t-1})$ obtained by Equation \eqref{eq:onestepahead} at time $t-1$ act like a "prior" distribution of $S^*_{t}$ given $\Omega_{t-1}$. On the other hand, this quantity is then corrected at time $t$ by the actual value of $Y_t$ through Equation \eqref{eq:filtprob}, resulting in the "posterior" distribution of $S^*_{t}$ given $\Omega_{t}$, $P(S^*_{t} = j\mid\Omega_{t})$. Both equations also play an important role for the inference for $S^*_{t}$, which translates to inference about the actual state $S_t$ at time $t$. This is subject of the following Section \ref{sec:smoothing}.\\

%(which describes the state-path from $t$ to $t-1$

% The filter probabilities $P(S^*_{t-1}\mid \Omega_{t-1})$ obtained by equation \eqref{eq:filtprob} at time $t-1$, summarize all the information $\Omega_{t-1}$ provides about $S^*_{t-1}$.

Last, the algorithm being recursive, it needs to be initialized. This part, carried out at $t=1$, can be completed by initializing only the equations \eqref{eq:lambda2} - \eqref{eq:onestepahead}, provided that we possess starting values for $Y_0$, $\Lambda_0$ and $P(S^*_{1} = j\mid\Omega_{0})$, $j = 1, \dots, m^2$. Starting values for $Y_0$ and $\Lambda_0$ and alternative initialization methods are discussed in detail in Appendix \ref{start}. Moreover, for both initializing the algorithm and throughout the optimization procedure we assume $(P(S^*_{1} = 1\mid\Omega_{0}), \dots, P(S^*_{1} = m^2\mid\Omega_{0})) = \delta^{*}$, where $\delta^{*}$ is the stationary distribution  of $S_{t}^*$. Appendix \ref{start} illustrates this part as well.  

% (See e.g. \cite{fruhwirth2006}).

%%%%%%%%%%%%%%%%%%%%%%%%%%%%%%%%%%%%%%%%%%%%%%%%%%%%%%%%%%%%%%%%%%%%%%%%%%%%%%%%%%%%%%%%%%
\subsection{Inference about the states}
\label{sec:smoothing}
%%%%%%%%%%%%%%%%%%%%%%%%%%%%%%%%%%%%%%%%%%%%%%%%%%%%%%%%%%%%%%%%%%%%%%%%%%%%%%%%%%%%%%%%%%

\begin{figure}[ht]
        \centering
\includegraphics[scale=1]{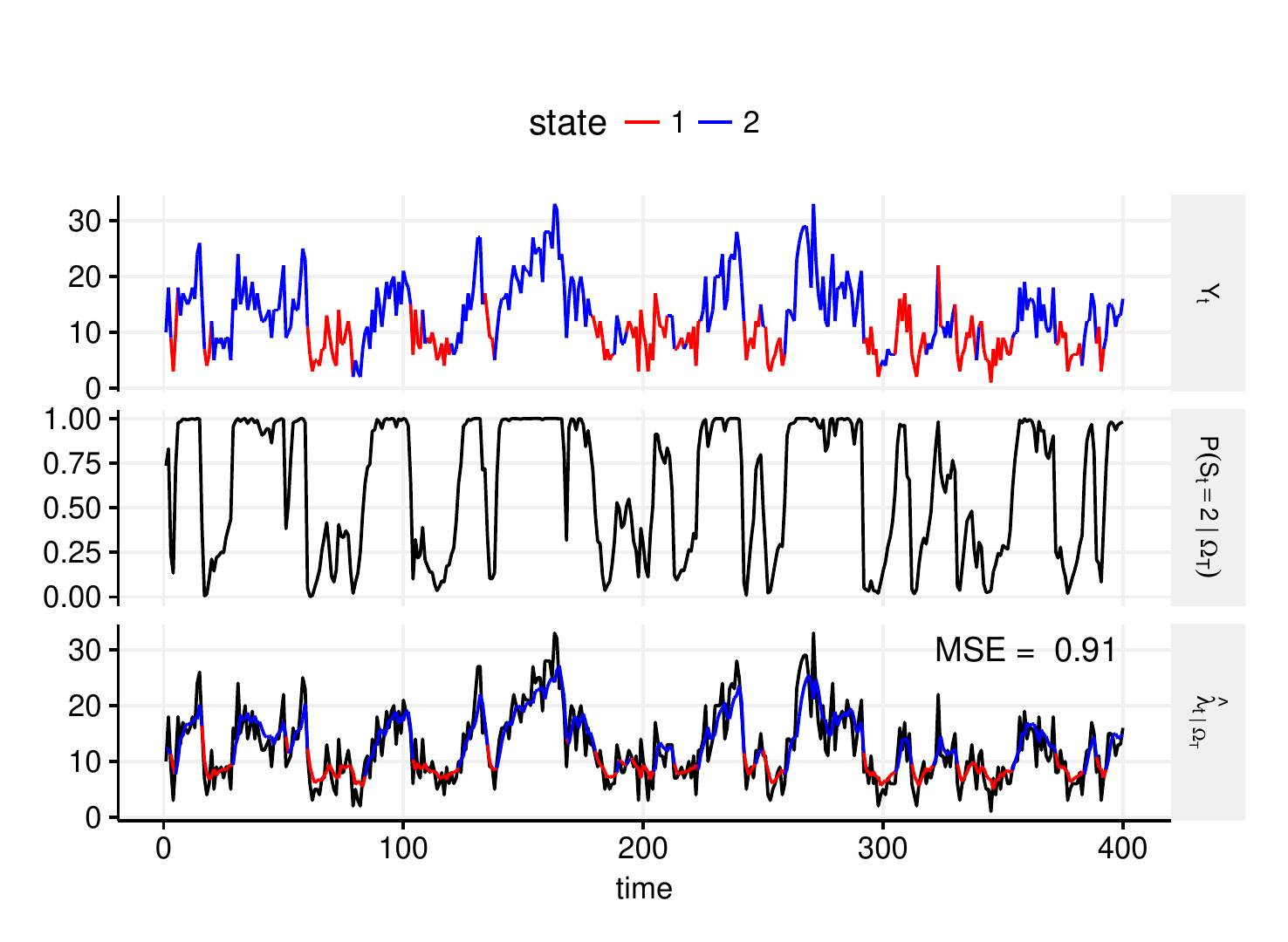}
       \caption{The top panel displays a simulated time series of length $T = 400$ from a $2$-state MS-PLLAR-model (Model of Case 2 in the simulation study). The coloring indicates the true state of the model. The middle panel displays the corresponding estimates of smoothing probabilities of being in state $2$, while the bottom panel displays predictions $\hat{\lambda}_{t|\Omega_{T}}$. The coloring in the bottom panel indicates the most probable state according to the smoothing probabilities.}
        \label{fig:simpred}
\end{figure}

Given the information set $\Omega_{\tau}$ with $\tau = \{1,\dots,T\}$, inference about the state $S_t$ at any time $t$ may be carried out via probabilities of the form $P(S_t = j \mid \Omega_{\tau})$, $j = 1,\dots,m$. Given analogous probabilities for the process $S^*_{t}$ defined by \eqref{Sstar}, $P(S_t = j \mid \Omega_{\tau})$ can be computed by   
% $P(S^*_{t} = j \mid \Omega_{\tau})$, $j = 1,\dots,m^2$
\begin{equation}
\label{eq:inf}
P(S_t = j \mid \Omega_{\tau}) = \sum_{i = 1}^{m^2} P(S_t = j, S^*_{t} = i \mid \Omega_{\tau}) = \sum_{i = 1}^{m^2} P(S^*_{t} = i \mid \Omega_{\tau}) \mathbbm{1}\left[S_t(S^*_{t} = i) = j\right],
\end{equation}
where $\mathbbm{1} \left[\cdot\right]$ corresponds to the indicator function. Thus, the filter probabilities $P(S_t = j \mid \Omega_t)$ and one-step ahead probabilities $P(S_{t+1} = j \mid \Omega_t)$, $j=1,\dots, m$, can be computed directly by \eqref{eq:inf} once the corresponding probabilities for the process $S^*_{t}$ have been obtained through the equations \eqref{eq:filtprob} and \eqref{eq:onestepahead}, respectively.\\
Furthermore, the smoothing probabilities $P(S_t\mid \Omega_T)$, $j = 1, \dots, m$, can be derived. These represent the inference about $S_t$ given the information set $\Omega_T$, and are of particular interest when analyzing data in in-sample settings. Similar to the filter and one-step ahead probabilities, the smoothing probabilities can be computed by \eqref{eq:inf}, provided that the corresponding smoothing probabilities for $S^*_{t}$ are available. For this purpose, we follow the approach of \cite{kim1994}. First, by the Markov property of $S^*_{t}$ we have that
\begin{equation*}
P(S^*_{t} = i\mid S^*_{t+1}  = j, \Omega_T)  = P(S^*_{t}  = i\mid S^*_{t+1} = j, \Omega_t)
 = \frac{\gamma^*_{ij}P(S^*_{t} = i\mid \Omega_t)}{P(S^*_{t+1} = j\mid\Omega_t)},
\end{equation*}  
 Secondly, the smoothing probabilities for $S^*_{t}$ can be represented as
\begin{equation}\label{eq:smoothprobs}
\begin{split}
P(S^*_{t} = i\mid \Omega_T) & = \sum_{j = 1}^{m^2}P(S^*_{t+1} = j\mid \Omega_T)P(S^*_{t} = i\mid S^*_{t+1}  = j, \Omega_T)\\
& = P(S^*_{t} = i\mid \Omega_t)\sum_{j = 1}^{m^2}\frac{\gamma^*_{ij}P(S^*_{t+1} = j\mid \Omega_T)}{P(S^*_{t+1} = j\mid\Omega_t)}
\end{split}
\end{equation} 
for $i,j = 1, \dots m^2$. Thirdly, using the quantities obtained from Equation \eqref{eq:filtprob} and \eqref{eq:onestepahead} as well as the filter (smoothing) probabilities $P(S^*_{T} = j\mid \Omega_T), j = 1, \dots, m^2$ as initial values, we are able to iterate backwards through Equation \eqref{eq:smoothprobs}. Hence, this recursive procedure permits to calculate the smoothing probabilities for $S_t$, $t = T - 1, \dots, 1$ via Equation \eqref{eq:inf}.\\
Estimates of the filter, one-step ahead, and smoothing probabilities result from replacing $\theta$ with the quasi maximum-likelihood estimate (QMLE) $\hat{\theta}$ in $\Omega_{\tau}$, where $\tau = t-1$, $t$ and $T$. Figure \ref{fig:simpred} provides an example for the estimated smoothing probabilities illustrated by means of a simulated time series.

%%%%%%%%%%%%%%%%%%%%%%%%%%%%%%%%%%%%%%%%%%%%%%%%%%%%%%%%%%%%%%%%%%%%%%%%%%%%%%%%%%%%%%%%%%
\subsection{Prediction and model assessment}
%%%%%%%%%%%%%%%%%%%%%%%%%%%%%%%%%%%%%%%%%%%%%%%%%%%%%%%%%%%%%%%%%%%%%%%%%%%%%%%%%%%%%%%%%%

A natural one-step ahead prediction $\hat{Y}_{T + 1}$ for $Y_{T + 1}$ is given by the expectation of $Y_{T+1}$ conditional on the information set $\Omega_T$, which includes information on potential covariate at time $T + 1$ by definition. Consistent with the Poisson assumption, denoting $\hat{Y}_{T + 1} = \hat{\lambda}_{T + 1\mid\Omega_{T}}$  follows  
\begin{equation}
\label{eq:pred1}
\begin{split}
\hat{\lambda}_{T + 1\mid\Omega_{T}} = E(y_{T + 1}\mid \Omega_T) &= \sum_{i = 1}^{m^2} E\left(y_{T + 1} \mid S^*_{T + 1} = i, \Omega_T\right)P(S^*_{T + 1} = i \mid \Omega_{T})\\
& = \sum_{i = 1}^{m^2} \hat{\lambda}_{T + 1\mid S^*_{T + 1} = i, \Omega_{T}}P(S^*_{T + 1} = i \mid \Omega_{T}), 
\end{split}
\end{equation}
where $P(S^*_{T + 1} = i \mid \Omega_{T})$ and $\hat{\lambda}_{{T + 1\mid S^*_{T + 1} = i, \Omega_{T}}} = \exp(\hat{\eta}_{T + 1\mid S^*_{T + 1} = i, \Omega_{T}})$ , $i = 1, \dots, m^2$ are available from the $T^{ \mbox{\tiny th}}$ and $T + 1^{\mbox{\tiny th}}$ recursion of the MS-PLLAR EHG algorithm, respectively. Provided that we possess covariate information up to time $T + k$, k-step ahead predictions $\hat{\lambda}_{T + k\mid\Omega_{T}}$ for $Y_{T + k}$ can also be obtained. For achieving this, the MS-PLLAR EHG algorithm needs to be executed up to time $T + k$ while iteratively replacing the unobserved observations $Y_{T + 1}, \dots, Y_{T + k - 1}$ in $\Omega_{T + 1}, \dots, \Omega_{T + k - 1}$ by their respective one-step ahead predictions $\hat{\lambda}_{T + 1\mid\Omega_{T}}, \dots, \hat{\lambda}_{T + k - 1\mid\Omega_{T}}$. In practice, $\theta$ is replaced by the QMLE $\hat{\theta}$ based on $y_1, \dots, y_T$ in $\Omega_{T + 1}, \dots, \Omega_{T + k - 1}$. Similarly, in case covariates are not observed beyond $T+1$, these also need to be replaced by some type of predicted values.

In a post-processing situation where $y_1, \dots, y_T$ have been observed, predictions of $y_1, \dots, y_T$ also provide valuable information about the model fit since they serve for computing residuals. Analogously to \eqref{eq:pred1}, one-step ahead predictions for $t = 1,\dots, T$ are given by
\begin{equation}\label{eq:pred2}
\begin{split}
\hat{\lambda}_{t\mid\Omega_{t - 1}} = E(y_{t}\mid \Omega_{t - 1}) &= \sum_{i = 1}^{m^2} E\left(y_{t} \mid S^*_{t} = i, \Omega_{t - 1}\right)P(S^*_{t} = i \mid \Omega_{t - 1})\\
& = \sum_{i = 1}^{m^2} \hat{\lambda}_{t\mid S^*_{t} = i, \Omega_{t - 1}}P(S^*_{t} = i \mid \Omega_{t - 1}), 
\end{split}
\end{equation}
where $P(S^*_{t} = i \mid \Omega_{t- 1})$ and $\hat{\lambda}_{{t\mid S^*_{t} = i, \Omega_{t - 1}}} = \exp(\hat{\eta}_{t\mid S^*_{t} = i, \Omega_{t - 1}})$, $i = 1, \dots, m^2$ are available from the $t - 1^{\mbox{\tiny th}}$ and $t^{\mbox{\tiny th}}$ recursion of the MS-PLLAR EHG algorithm, respectively. However, this prediction can be improved by utilizing the smoothing probabilities, which are available in a post-processing situation, thus
\begin{equation}
\label{eq:pred3}
\hat{\lambda}_{t\mid\Omega_{T}} = \sum_{i = 1}^{m^2} \hat{\lambda}_{t\mid S^*_{t} = i, \Omega_{t - 1}}P(S^*_{t} = i \mid \Omega_{T}), 
\end{equation}
where the notation $\hat{\lambda}_{t\mid\Omega_{T}}$ is somewhat lax since $\hat{\lambda}_{t\mid\Omega_{T}} \neq E(Y_t \mid \Omega_T)$. Figure \ref{fig:simpred} displays this prediction method for a simulated time series. An alternative approach is to recursively compute $\eta_t$ in \eqref{eq:msloglinear} along the regime-path deemed most likely by the smoothing probabilities and take $\exp(\eta_t)$ as an in-sample predictor of $Y_t$. However, this will be a poor prediction if the states are not well separated, or if smooth transition periods between regimes occur in the data. Hence, in general weighted averages such as \eqref{eq:pred2} and \eqref{eq:pred3} are preferred, and we will use \eqref{eq:pred3} in what follows. Note that the plug-in value of $\theta$ in the MS-PLLAR EHG algorithm is the only difference between in-sample and out-of-sample predictions.\\ %For the former, $\theta$ is estimated based on $y_1, \dots, y_T$, whereas it is based on a training data set for the latter. %, however the model is applied to the observations $y_1, \dots, y_T$. 
Given predictions of $y_1,\dots, y_T$, we can compute the Pearson residuals by
\begin{equation}
r_t = (y_t - \hat{\lambda}_{t\mid \Omega_{T}})/\sqrt{\hat{\lambda}_{t\mid \Omega_{T}}}
\end{equation}
for $t = 1,\dots, T$. Under the correct model, the sequence $r_t$ should resemble white noise with constant variance. The empirical autocorrelation function (ACF) of these residuals can be inspected to check for the presence of serial dependence which is not captured by the model. Following \citep[][section 1.6 and 1.8]{2005kedem}, the mean square error (MSE) of the Pearson residuals given by $\sum_{t = 1}^T r_{t}^2/(T - p)$, where $p$ denotes the number of parameters in the model, serves for evaluating competing models. Last, the Poisson assumption can be inspected by plotting the predictions $\hat{\lambda}_{t\mid \Omega_{T}}$ against the squared raw residuals $(y_t - \hat{\lambda}_{t\mid \Omega_{T}})^2$. In this plot, points scattering symmetrically around the line $y = x$ indicated a good model fit.

\section{Simulation and empirical analysis}
%%%%%%%%%%%%%%%%%%%%%%%%%%%%%%%%%%%%%%%%%%%%%%%%%%%%%%%%%%%%%%%%%%%%%%%%%%%%%%%%%%%%%%%%%%

In this section we present results of a simulation study and an empirical analysis corporate defaults.

%%%%%%%%%%%%%%%%%%%%%%%%%%%%%%%%%%%%%%%%%%%%%%%%%%%%%%%%%%%%%%%%%%%%%%%%%%%%%%%%%%%%%%%%%%
\subsection{Simulation study}
\label{sec:sim}
%%%%%%%%%%%%%%%%%%%%%%%%%%%%%%%%%%%%%%%%%%%%%%%%%%%%%%%%%%%%%%%%%%%%%%%%%%%%%%%%%%%%%%%%%%

In the following, we report the results from a simulation study designed for assessing the finite sample performance of the QMLE's derived in Section \ref{sec:ehg_algo}. The study bases on $1000$ simulated time series of length $T=200, 500, 1000$, respectively, from two-state MS-PLLAR models subject to different parametrizations. These are termed Case 1 and Case 2, Table \ref{table:cases} summarizes the different parameter values: in Case 1 the two regimes are well separated in terms of both dependence structure (parameter $a$ and $b$, respectively) and level ($d$ parameter, relative to $a$ and $b$). The parameters of the first regime are taken from the simulation study conducted in \cite{fokianos2011}, and produce a time series with negative correlations at lag one. On the contrary, time series with strong positive correlation for several lags are characteristic for the second regime. Moreover, averaging the time series value in regime one for very long simulated time series results in the value $1.30$, compared to $15.64$ in regime two. For Case 2, the differences between regimes are more subtle. Both regimes produces positive correlations for several lags, but with stronger lag correlations in the second than in the first regime.  The long run average in state one equals $8.24$, compared to $15.64$ in state two. Therefore, compared to Case 2 one can expect higher precision of the estimates in Case 1. Figure \ref{fig:simpred} displays a simulated time series for Case 2. The true parameter values of each case served for initializing the estimation procedure.\\   

\begin{figure}[ht]
        \centering
\includegraphics[scale=.8]{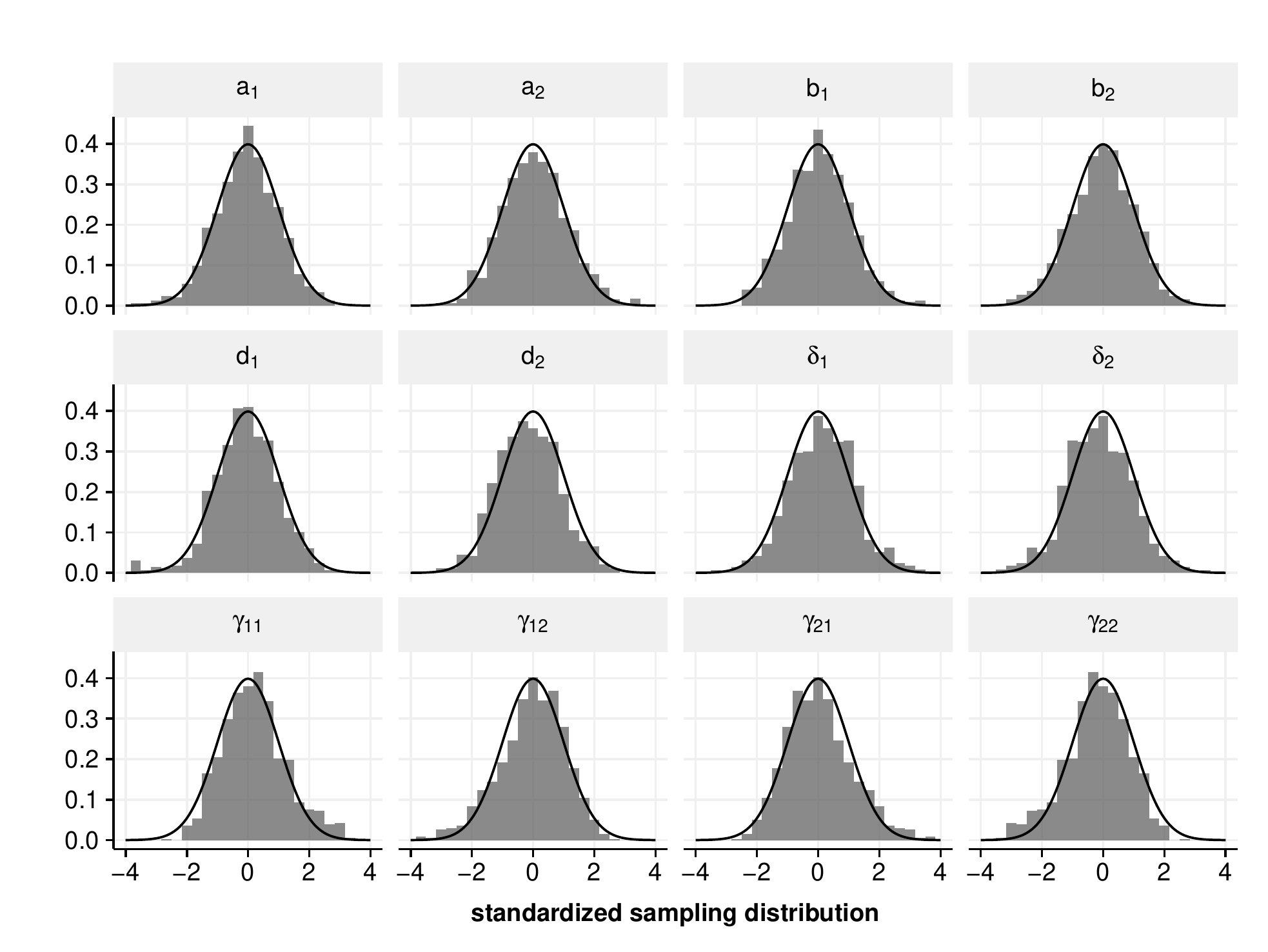}
       \caption{Relative frequency of the standardized quantities $(\hat{\theta} - \theta_0)/se(\hat{\theta})$ obtained from each run compared to the standard 
       normal density. Case 1, $T = 500$.}
        \label{fig:sim_asym1}
\end{figure}

\begin{figure}[ht]
        \centering
\includegraphics[scale=.8]{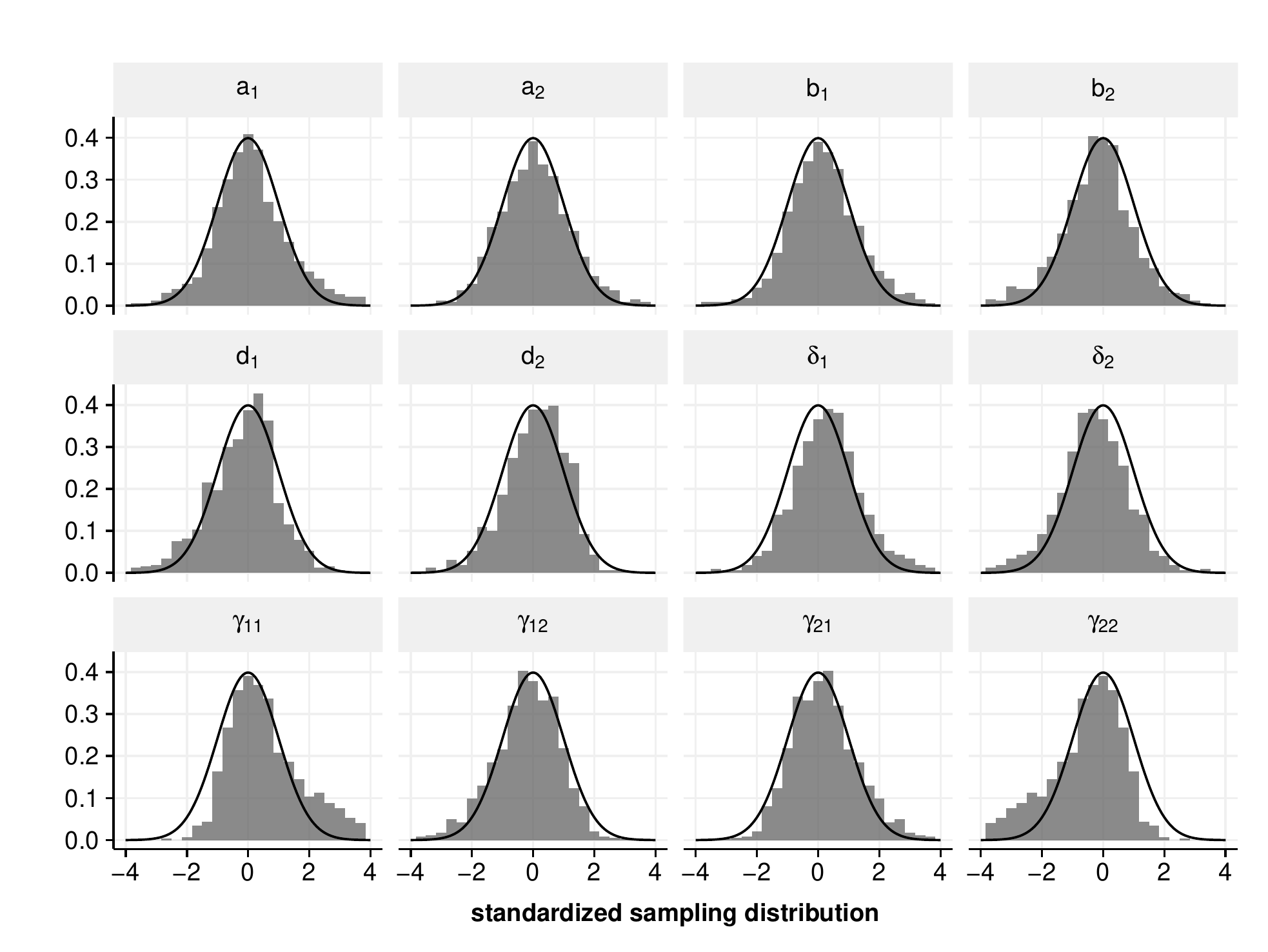}
       \caption{Relative frequency of the standardized quantities $(\hat{\theta} - \theta_0)/se(\hat{\theta})$ obtained from each run compared to the standard 
       normal density. Case 2, $T = 500$.}
        \label{fig:sim_asym2}
\end{figure}

Table \ref{table:sim} summarizes the results of the simulation study. The bias values correspond to the average estimated value of all runs minus the corresponding true parameter value. Similarly, the standard error (SE) is defined as the sample standard deviations of the estimates obtain by simulation. We also investigate the adequacy of the standard error $se(\hat{\theta})$ described in Appendix \ref{sec:implementation details}, which is based on the delta-method and the exact Hessian. For this purpose, the reports the average estimated standard error of all runs ($\widehat{\text{SE}}$) as well, which can be compared in turn with the sample standard deviation (SE).\\
With the exception of Case 2 with $n=200$, the bias is low. In both Case 1 and Case 2 the SE decreases as $n$ increases, but as expected there is more uncertainty related to the parameters in the second case. In particular for $n=200$, the standard error ($\widehat{\text{SE}}$) seems to be slightly underestimating compared to the sample standard deviations (SE), which is not atypical for models of such complexity. However, for $n = 500, 1000$ SE and $\widehat{\text{SE}}$ are approach each other. Figure \ref{fig:sim_asym1} and \ref{fig:sim_asym2} display the relative frequency of the standardized quantities $(\hat{\theta} - \theta_0)/se(\hat{\theta})$ obtained from each run compared to the standard normal density for the two cases.  Apart from the parameters of the Markov chain, which lie close to the border of the set of possible values, all other parameters show not stronger deviations from normality.

%%%%%%%%%%%%%%%%%%%%%%%%%%%%%%%%%%%%%%%%%%%%%%%%%%%%%%%%%%%%%%%%%%%%%%%%%%%%%%%%%%%%%%%%%%
\subsection{Empirical analysis}
%%%%%%%%%%%%%%%%%%%%%%%%%%%%%%%%%%%%%%%%%%%%%%%%%%%%%%%%%%%%%%%%%%%%%%%%%%%%%%%%%%%%%%%%%%

\begin{figure}[ht]
        \centering
\includegraphics[scale=0.8]{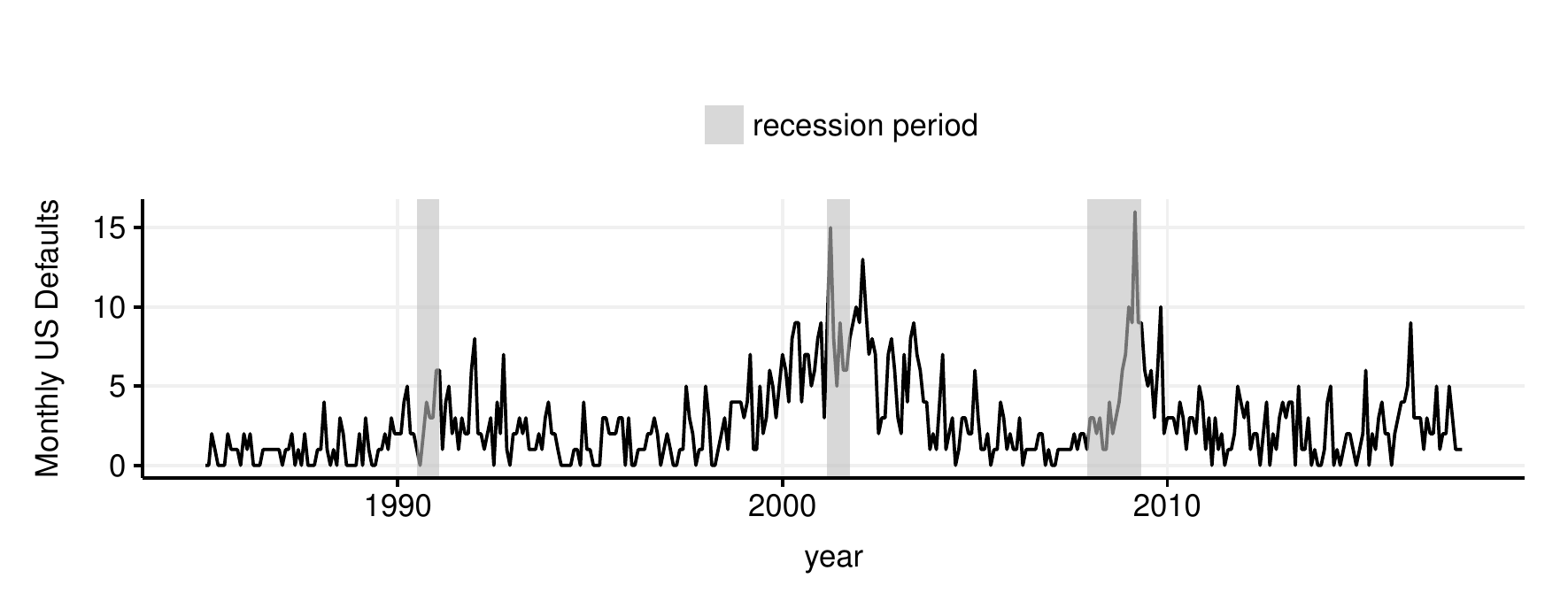}
       \caption{Monthly data from January 1985 to September 2017.}
        \label{fig:data}
\end{figure}

\begin{figure}[ht]
        \centering
\includegraphics[scale=0.8]{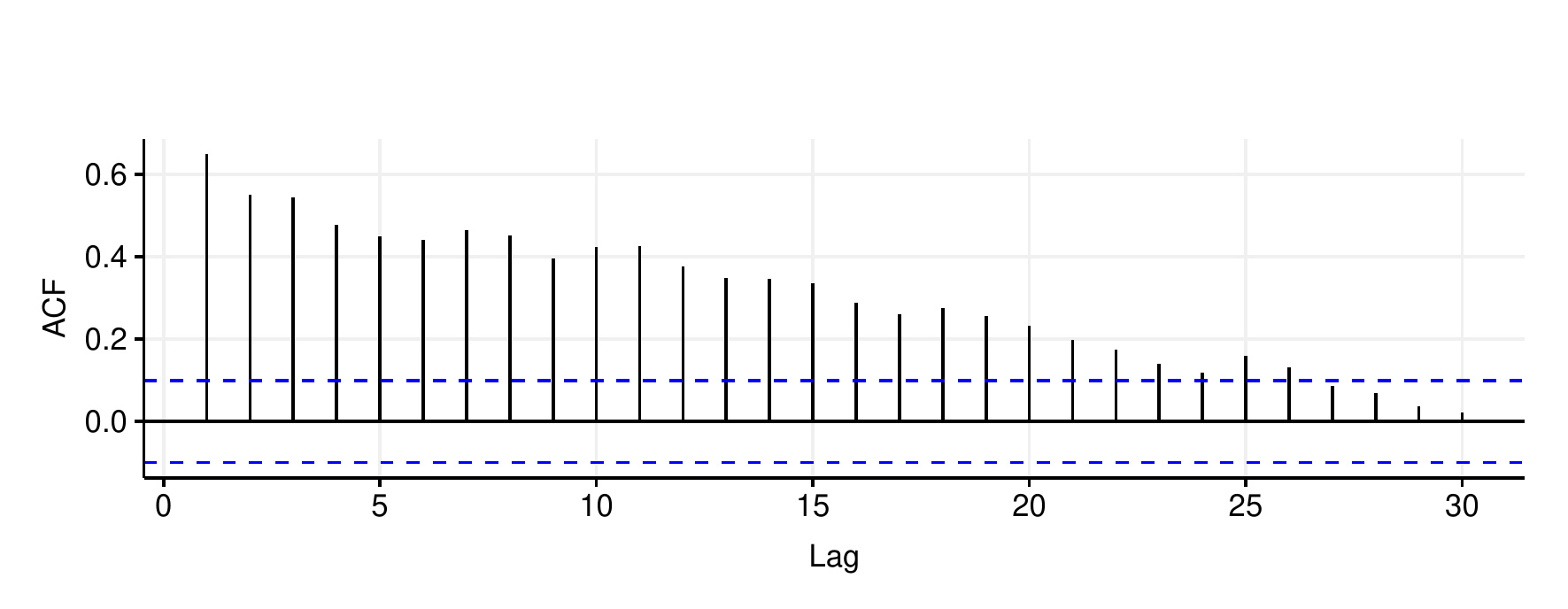}
       \caption{The autocorrelation function of the monthly number of defaults}
        \label{fig:acf}
\end{figure}

In this section we provide an analysis of corporate default counts in the US, using the MS-PLLAR model introduced in Section 2. The US defaults count data corresponds to the monthly number of bankruptcies filed in the United States Bankruptcy courts, and is available from the UCLA-LopPucki Bankruptcy Research database (see \emph{http://lopucki.law.ucla.edu}). These data cover the period from January 1985 to September 2017, in total 393 monthly observations. It consists of the counts of defaults of all large, public companies, where large is defined as having declared more than US\$ 100 million in assets the year before the firm filed the bankruptcy case, measured in 1980 dollars. A company is considered public if it had reported to the Securities and Exchange Commission (SEC) in the last three years prior to the bankruptcy. The count of monthly bankruptcies are aggregated by the calender month in which the bankruptcy was filed. Over the sample period a total of 1065 defaults is counted, Figure \ref{fig:data} displays the time series together with recession periods. The recession periods used, are the NBER based recession indicators for the United States (USREC) available from the St. Louis Fed online database FRED.
% {\color{blue} (Maybe include recessions in grey here? or at least in some other plots) BARD/GEIR: doe we have an objective quantification of recession periods? I used something like this in a paper of mine some time ago, but am not sure if its consistent with finance papers / definitions used by santanna if he does. if no time now, please put it on the todo list}. 
 Figure \ref{fig:acf} shows a plot of the autocorrelation function of the observations. As highlighted by other studies, these two figures illustate some stylized facts: first, the existence of default clusters; second, the high temporal dependence in the count of defaults; third, overdispersion of the distribution of default counts, as the empirical average is 2.42 while the empirical variance is 6.50. Even though the default counts are available since October 1979, we only use data from 1985 onward to avoid some extreme structural breaks in the covariates, cfr.~\cite{santanna2017}. %{\color{blue} GEIR: can you check for the right value? BS: The number of observations is already given in the beginning of this section}\\ 
These data have already been studied by several other authors \citep[e.g.][]{santanna2017}, covering a slightly shorter time span. Furthermore, other studies \citep[e.g.][]{agosto2016, azizpour2017} base on data exhibiting comparable dynamic patterns from Moody's Default Risk Service.\\ % and does not appear significantly different by visual inspection.  

The purpose of the study is to examine whether the common systematic risk variables can explain the default clustering observed, or if there is default clustering beyond this, i.e.~due to the presence of a contagion effect. In addition, as we fit Markov-switching models, we are able to examine whether the effect of the covariates are time-heterogeneous or not. Finally, we are able to reveal if the contagion effect is present in all regimes or not.

%%%%%%%%%%%%%%%%%%%%%%%%%%%%%%%%%%%%%%%%%%%%%%%%%%%%%%%%%%%%%%%%%%%%%%%%%%%%%%%%%%%%%%%%%%
\subsection{Excluding exogenous covariates}
%%%%%%%%%%%%%%%%%%%%%%%%%%%%%%%%%%%%%%%%%%%%%%%%%%%%%%%%%%%%%%%%%%%%%%%%%%%%%%%%%%%%%%%%%%

We start the empirical analysis by excluding covariates, and focus on determining the number of regimes present for the corporate default series. That is, we fit model \eqref{eq:msloglinear} excluding the term $\beta^\prime X_t$, and set $m$ equal to  1, 2, and 3 regimes. Table \ref{table:comp1} reports a comparison between the different models, while Table \ref{table:pars} shows the estimated parameters for the three models. The MSE marginally favors the model with three regimes, however both the AIC and BIC rank the model with two regimes above the model with $m=3$. The one-state model is ranked last, except when using the BIC. Hence, the two-state model represents a suitable choice overall.\\

We further note that the parameter estimates for the model with one regime correspond relatively well to those obtained for the second state in the model with $m=2$ (i.e.~$a_2, b_2$ and $d_2$ are comparable to $a, b$ and $d$). The positive sign of the estimated $b$ parameter indicates that the previously observed number of defaults increases the intensity in the current month. However, we cannot reject that $d=0$, $a=1$, $b=0$ for the first state in the two-state model. Consequently, solving $\eta = d + a\eta + b \log(y_{t-1} + 1)$ for $\eta$ results in $\eta = \eta$, indicating a constant intensity. In other words, this model can be characterized by one regime with close to constant default intensity one the one hand, and a second state subjet to more dynamics on the other hand.\\

\begin{figure}
        \centering
\includegraphics[scale=0.8]{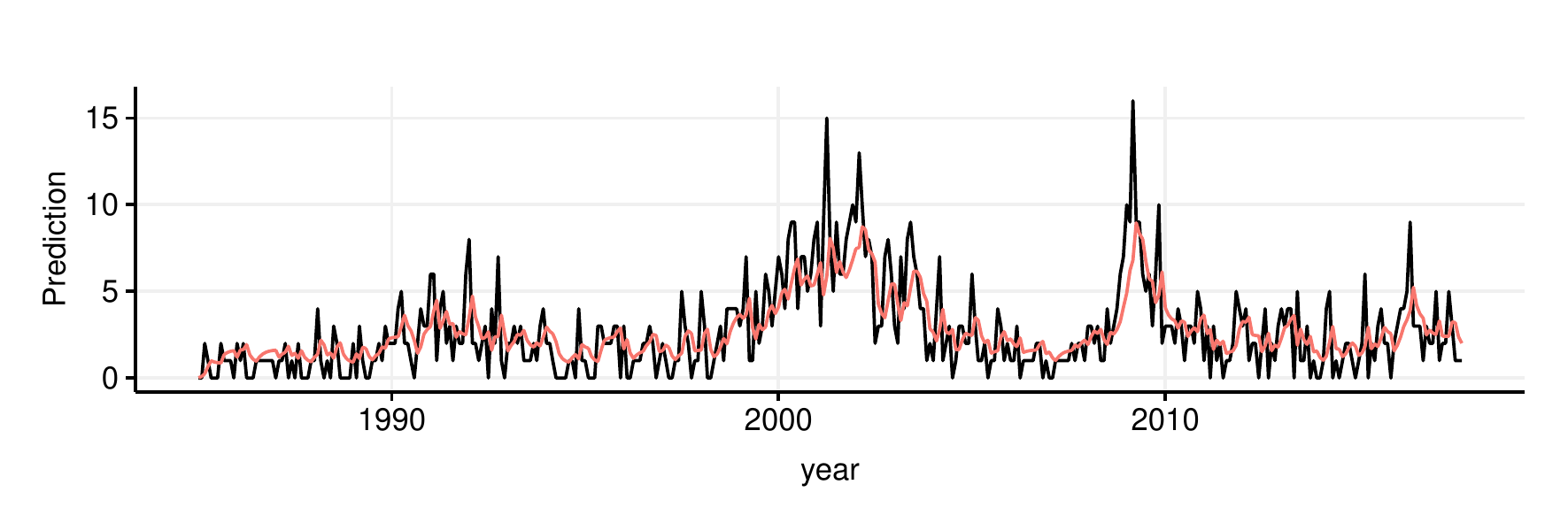}
       \caption{Prediction when $m = 1$}
        \label{fig:mod1}
\end{figure}

\begin{figure}[ht]
        \centering
\includegraphics[scale=0.8]{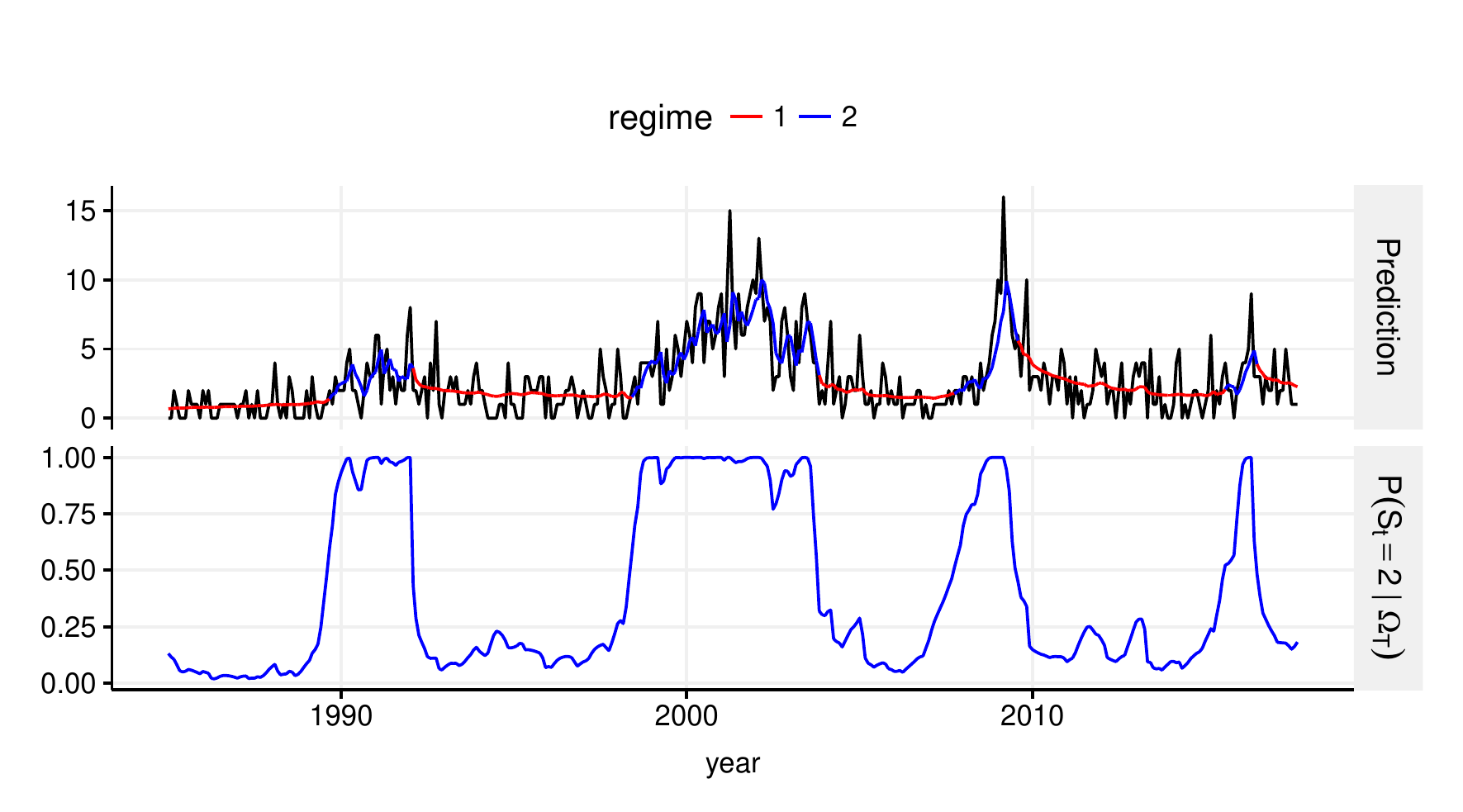}
       \caption{Prediction and smoothing probabilities for $m = 2$}
        \label{fig:mod2}
\end{figure}

\begin{figure}[ht]
        \centering
\includegraphics[scale=0.8]{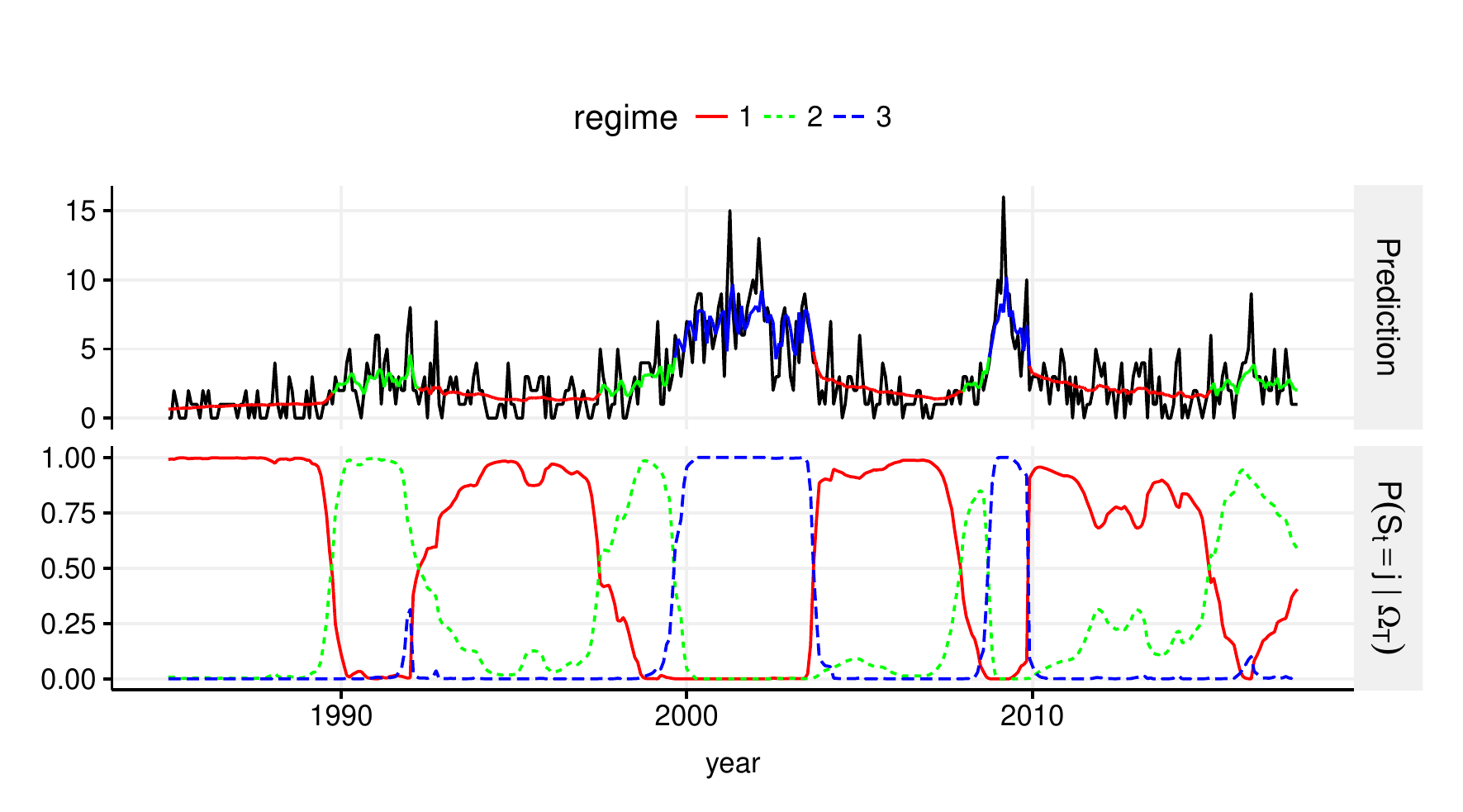}
       \caption{Prediction and smoothing probabilities for $m = 3$}
        \label{fig:mod3}
\end{figure}

The model with three regimes resemble the model with two regimes, but with an additional "medium" dynamic state, as seen from the parameter estimates. Figures \ref{fig:mod1}, \ref{fig:mod2}, and \ref{fig:mod3} show predictions from the fitted models, and the smoothing probabilities for the model with $m$ equal to two and three, respectively. Based on this analysis and the model comparison, we remain with our previous conclusion that two-regime model is a suitable approach for extending our analysis by including covariates in the intensity equation. 

%Comment state 1 for $m=2$:

%%%%%%%%%%%%%%%%%%%%%%%%%%%%%%%%%%%%%%%%%%%%%%%%%%%%%%%%%%%%%%%%%%%%%%%%%%%%%%%%%%%%%%%%%%
\subsection{Including exogeneous covariates}
%%%%%%%%%%%%%%%%%%%%%%%%%%%%%%%%%%%%%%%%%%%%%%%%%%%%%%%%%%%%%%%%%%%%%%%%%%%%%%%%%%%%%%%%%%

We will use a number of macroeconomic and financial variables that represent the common systematic risk corporations face %BARD: is this a common term? BS: Yes 
as explanatory variables. Similar to \cite{santanna2017}, we use monthly variables collected from the St. Louis Fed online database FRED. The variables considered are the industrial production index (INDPRO), new housing permits (PERMIT), civilian unemployment rate (UNRATE), Moody's seasoned baa corporate bond yield (BAA), 10-years treasury constant maturity rate (GS10), federal funds rate (FEDFUNDS), producer price index by commodity for final demand: finished goods (PPIFGS), and produce price index: fuels and related energy (PPIENG). In addition, we collected the variables S{\&}P500 annualized returns (SP500ret) and S{\&}P500 annualized return volatility (SP500vol) from DataStream.\\ 
The variables INDPRO, PERMIT, PPIFGS and PPIENG are expressed as yearly growth rates, whereas the variables UNRATE, BAA, FEDFUNDS, GS10, SP500ret and SP500vol are expressed as yearly differences. %Table \ref{tab1} display some basic summary statistics of all variables. - Consider this in the submission version. 
 Most of these covariates have been found to have significant impact on default rates and were used in similar studies  \cite[see, e.g.,][]{das2007, duffie2009, giesecke2011, agosto2016, azizpour2017}.\\

\begin{figure}[ht]
        \centering
\includegraphics[scale=0.8]{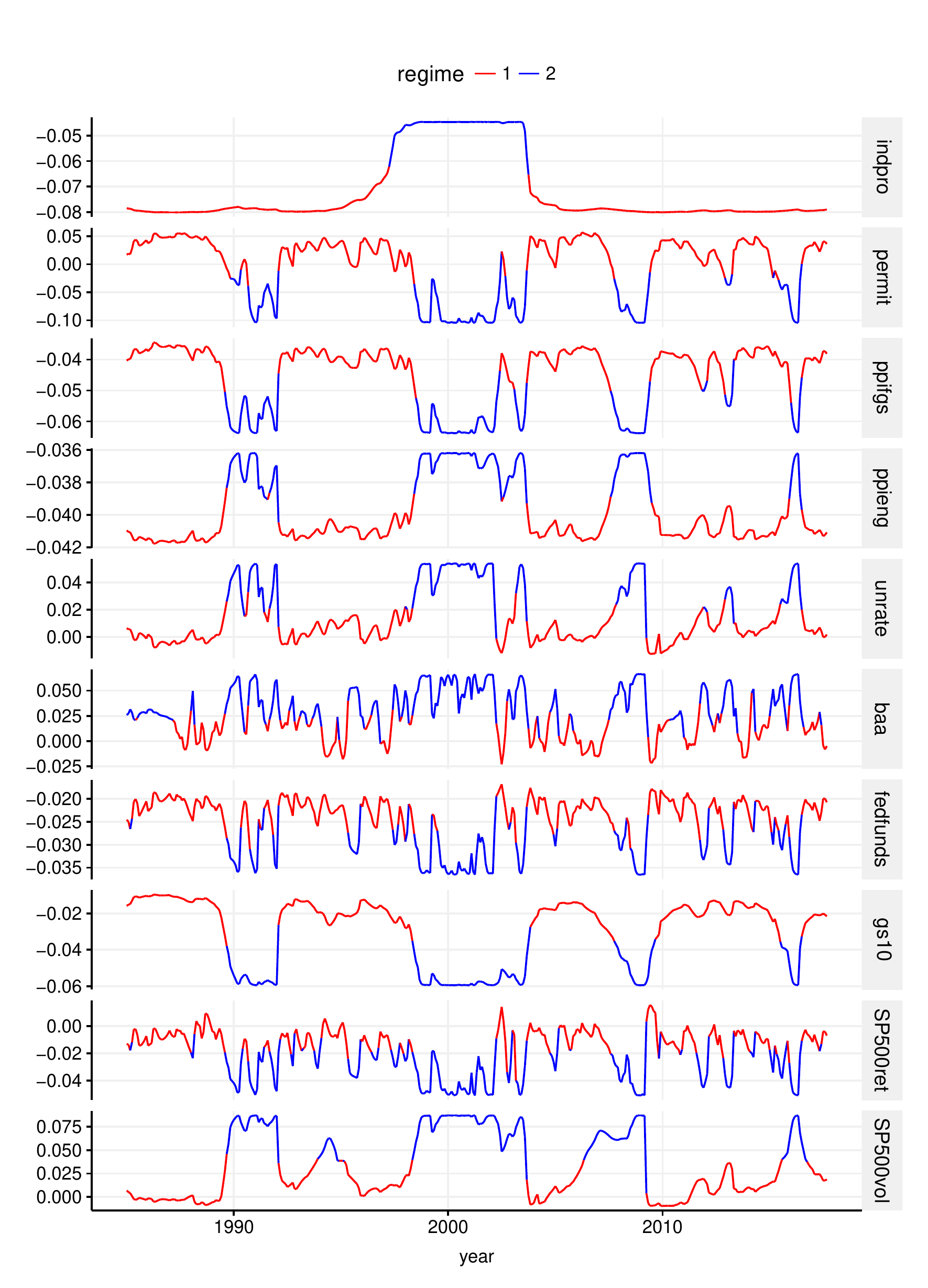}
       \caption{Temporal trajectories of the covariate effects. The coloring indicates the most probable state according to the smoothing probabilities for the model including the respective covariate.} 
        \label{fig:smoothcov}
\end{figure}

As described in the section above, we apply model \eqref{eq:msloglinear} with two regimes, and fit separate models using only one covariate for each. This results in ten fitted models, Table \ref{table:cov1} reports a model comparison. In the same table, we also report whether the covariate included in each model is found significant or not in any of the two regimes. A clear pattern occurring is that none of the covariates is significant in both regimes, and most are only significant in the most dynamic regime (i.e.~number two). In particular, the covariates related to the financial market (SP500ret, SP500vol) are significant in the second regime, which is in line with findings of \cite{agosto2016}. Figure \ref{fig:smoothcov} displays the temporal trajectories of the covariate effects obtained from
\begin{equation*}
\hat{\beta}(t) = \hat{\beta}_1 P(S_t = 1 \mid \Omega_{T}) + \hat{\beta}_2 P(S_t = 2 \mid \Omega_{T}) 
\end{equation*}
These trajectories indicate the temporal variation of covariate effects on the number of defaults.\\

As noted in Section \ref{sec:interpretation}, the parameter $b$ should be equal to zero in the case of conditional independence. From the estimates of this parameter, we test the null hypothesis $H_0: b_m=0$ (for $m=1$ and $2$, i.e. in both regimes separately). The results show that this hypothesis is rejected for all models for the second regime, but cannot be rejected for all models for the first regime. This implies the presence of contagion in the second regime, but not in the first, thus the notion of contagion is indeed time-varying. These findings are in line with \cite{agosto2016}, where systematic risk factors have been able to explain the default clustering observed in the recent years by a more ad-hoc approach of fitting models to sampling periods lying in different time windows.

%%%%%%%%%%%%%%%%%%%%%%%%%%%%%%%%%%%%%%%%%%%%%%%%%%%%%%%%%%%%%%%%%%%%%%%%%%%%%%%%%%%%%%%%%%
\section{Concluding remarks and outlook}
%%%%%%%%%%%%%%%%%%%%%%%%%%%%%%%%%%%%%%%%%%%%%%%%%%%%%%%%%%%%%%%%%%%%%%%%%%%%%%%%%%%%%%%%%%

In this paper, we have introduced the Markov-switching Poisson log-linear autoregressive (MS-PLLAR) model, and developed a QMLE using an adaptation of the extended Hamilton-Grey (EHG) algorithm to avoid path-dependence problems. A simulation study indicates that the proposed QMLE is well-behaved. The MS-PLLAR model is suitable to model count time series of corporate defaults, as they are correlated over time and exhibit the default clustering effect, i.e.~high peaks in clusters.\\
By using the MS-PLLAR model, we provide evidence that the time series of counts of US default consist of two regimes and that the contagion effect, i.e. that past defaults impact the probability that firms default in the future, is present in one of these regimes. We also note that the coefficients of the covariates are different in each of the regimes. In conclusion, the notion of contagion in the default process is slightly more delicate than previously believed.\\

In the paper, we have only fitted models with one covariate. Thus, the natural next step in the empirical analysis is to include the most significant covariates successively in the model, and then perform the the test for contagion as above. We leave this for future research. Moreover, several alternative model specifications come to mind as potential research subjects as well. For example, the inclusion of further lags for covariates and response or different link functions. % alternative inclusion of covariates; Not truly exogenous now??
In addition, other choices of conditional distribution $Y_t\mid \mathcal{F}_{t - 1}$ are possible. For example, one can assume $Y_t\mid \mathcal{F}_{t-1}\sim NegBin(\lambda_{t,k}, \phi_k)$ where \citep[see][]{christou2014} the negative binomial distribution is parameterized in terms of its (state-dependent) mean $\lambda_t$ and a (state-dependent) dispersion parameter $\phi_{S_t}$:
\begin{equation}
P(Y_{t} = y\mid \mathcal{F}_{t-1}) = \frac{\Gamma(\phi_{S_t} + y)}{\Gamma(y + 1)\Gamma(\phi_{S_t})}\left(\frac{\phi_{S_t}}{\phi_{S_t} + \lambda_t}\right)^{\phi_{S_t}}\left(\frac{\lambda_t}{\phi_{S_t} + \lambda_t}\right)^y
\end{equation}
It follows that $\Var(Y_{t} = y\mid \mathcal{F}_{t-1}) = \lambda_{S_t} + \lambda_t^2/\phi_{S_t}$ in contrast to the Poisson case where $\Var(Y_{t} = y\mid \mathcal{F}_{t-1}) = \E(Y_{t} = y\mid \mathcal{F}_{t-1}) = \lambda_t$. The mean parameter $\lambda_t$ can be modeled both with  a linear and log-linear conditional mean. Such an extension should not pose major obstacles, since, on the one hand, the estimation procedure described in Section \ref{sec:estimation} is not confined to the Poisson distribution nor the log-linear specification of the conditional mean given in \eqref{eq:msloglinear}. On the other hand, however, some modifications are needed to accommodate regression on past values $\eta_{t - l}$ and $Y_{t - l}$ for $t > 1$. These modifications entails tracing the state-paths over more lags and expanding the information set \eqref{eq:iset} analogously to the procedure described in \cite{chen2011}. 

\clearpage %Before bibliography

\bibliographystyle{agsm} %plainnat
\bibliography{ms}

\addcontentsline{toc}{chapter}{References}

\clearpage

%%%%%%%%%%%%%%%%%%%%%%%%%%%%%%%%%%%%%%%%%%%%%%%%%%%%%%%%%%%%%%%%%%%%%%
%% Figures
%%%%%%%%%%%%%%%%%%%%%%%%%%%%%%%%%%%%%%%%%%%%%%%%%%%%%%%%%%%%%%%%%%%%%%

\pagebreak

% % Figure of one step ahead prediction and smoothing probs
% \begin{figure}[h]
%         \centering
% \includegraphics[scale=1]{figures/sim_states.pdf}
%        \caption{Top: A simulated timeseries of length $T = 400$ from a $2$-state INGARCH-model with parameters $\theta = \left((d_1, d_2) = (1, 0.3), (a_1, a_2) = (0.2, 0.4), (b_1, b_2) = (0.3, 0.5), (\gamma_{11}, \gamma_{12}, \gamma_{21}, \gamma_{22}) =  (0.90, 0.10, 0.10, 0.90)\right)$. The colouring indicates the true state of the model. Mid and bottom figure displays the corresponding estimates of prediction and smoothing probabilities of beeing in state $2$, respectively.}
%         \label{fig:simstates}
% \end{figure}
% 
% % Figure of predictions
% \begin{figure}[h]
%         \centering
% \includegraphics[scale=1]{figures/sim_pred.pdf}
%        \caption{In-sample predictions for the timeseries displayed in figure \ref{fig:simstates}. The top panel displays the one-step ahead prediction $\hat{\lambda}_{t|\Omega_{t-1}}$, the middle panel $\exp(\eta_t)$ in \eqref{eq:msloglinear} recursively computed along the regime-path deemed most likely by the smoothing probabilities, and the bottom panel $\hat{\lambda}_{t|\Omega_{T}}$ defined by \eqref{eq:pred3}. The colouring is based on the most probable state according to the prediction probabilities $P(S_t = t\mid \Omega_{t-1})$ (top) and smoothing probabilities $P(S_t = t\mid \Omega_{t-1})$ (mid and bottom). Note that the preference of $\hat{\lambda}_{t|\Omega_{T}}$ is supported by the MSE.}
%         \label{fig:simpred}
% \end{figure}

% Figure of both smoothing probs and state predictions

% Figures of simulation study

\pagebreak

\begin{table}[ht]
\caption{Overview of the two cases of parameter values}
\centering
\scalebox{0.8}{
% \begin{tabular}{l l@{\hspace{.5cm}} c c c c@{\hspace{.5cm}} c c c c@{\hspace{.5cm}} c c c c}
\begin{tabular}{l c c c c c c c c c c}
\toprule
%\multicolumn{1}{c}{} & \multicolumn{3}{l}{Regime 1} & \multicolumn{3}{l}{Regime 2} & \multicolumn{4}{l}{$\Gamma$} &
& \multicolumn{3}{l}{Regime 1} & \multicolumn{3}{l}{Regime 1} & \multicolumn{4}{l}{$\Gamma$}\\
\cmidrule(lr){2-4}\cmidrule(lr){5-7}\cmidrule(lr){8-11}
& $a_1$ & $b_1$ & $d_1$ & $a_2$ & $b_2$ & $d_2$ & $\gamma_{11}$ & $\gamma_{21}$ & $\gamma_{12}$ & $\gamma_{22}$\\
\midrule
Case 1 & -0.5 & -0.35 & 0.50 & 0.40 & 0.50 & 0.30 & 0.95 & 0.05 & 0.05 & 0.95\\
  \noalign{\vskip 3mm}
Case 2 & 0.20 & 0.30 & 1.00 & 0.40 & 0.50 & 0.30 & 0.90 & 0.10 & 0.10 & 0.90\\ 
  \bottomrule
\end{tabular}
}
\label{table:cases}
\end{table}

\begin{table}[ht]
\caption{Result of simulation study}
\centering
\scalebox{0.8}{
% \begin{tabular}{l l@{\hspace{.5cm}} c c c c@{\hspace{.5cm}} c c c c@{\hspace{.5cm}} c c c c}
\begin{tabular}{l l c c c c c c c c}
\toprule
& & \multicolumn{4}{l}{Case 1}& \multicolumn{4}{l}{Case 2}\\
\cmidrule(lr){3-6}\cmidrule(lr){7-10}  
Sample size&Parameter & Value & Bias & $\text{SE}$ & $\widehat{\text{SE}}$ & Value & Bias & $\text{SE}$ & $\widehat{\text{SE}}$\\
\midrule
$200$ & $a_1$ & -0.50 & 0.0142 & 0.1322 & 0.1223 & 0.20 & -0.0358 & 0.2611 & 0.2048 \\ 
  &$a_2$ & 0.40 & 0.0151 & 0.1328 & 0.1102 & 0.40 & 0.0091 & 0.1539 & 0.1367 \\ 
  &$b_1$ & -0.35 & -0.0127 & 0.2042 & 0.1865 & 0.30 & -0.0091 & 0.1439 & 0.1283 \\ 
  &$b_2$ & 0.50 & -0.0167 & 0.1275 & 0.1110 & 0.50 & -0.0331 & 0.1298 & 0.1190 \\ 
  &$d_1$ & 0.50 & -0.0070 & 0.1533 & 0.1510 & 1.00 & 0.0917 & 0.4761 & 0.3748 \\ 
  &$d_2$ & 0.30 & 0.0028 & 0.0987 & 0.0960 & 0.30 & 0.0680 & 0.2268 & 0.2076 \\ 
  &$\gamma_{11}$ & 0.95 & -0.0018 & 0.0253 & 0.0233 & 0.90 & -0.0059 & 0.0881 & 0.0663 \\ 
  &$\gamma_{21}$ & 0.05 & 0.0064 & 0.0286 & 0.0244 & 0.10 & 0.0016 & 0.0569 & 0.0498 \\ 
  &$\gamma_{12}$ & 0.05 & 0.0018 & 0.0253 & 0.0233 & 0.10 & 0.0059 & 0.0881 & 0.0663 \\ 
  &$\gamma_{22}$ & 0.95 & -0.0064 & 0.0286 & 0.0244 & 0.90 & -0.0016 & 0.0569 & 0.0498 \\ 
  &$\delta_{1}$ & 0.50 & 0.0200 & 0.1256 & 0.1251 & 0.50 & 0.0115 & 0.1478 & 0.1458 \\ 
  &$\delta_{2}$ & 0.50 & -0.0200 & 0.1256 & 0.1251 & 0.50 & -0.0115 & 0.1478 & 0.1458 \\ 
  \noalign{\vskip 3mm} 
$500$ & $a_1$ & -0.50 & 0.0005 & 0.0743 & 0.0726 & 0.20 & -0.0036 & 0.1508 & 0.1413 \\ 
  &$a_2$ & 0.40 & 0.0029 & 0.0708 & 0.0654 & 0.40 & 0.0004 & 0.0968 & 0.0911 \\ 
  &$b_1$ & -0.35 & 0.0011 & 0.1135 & 0.1117 & 0.30 & 0.0060 & 0.0913 & 0.0864 \\ 
  &$b_2$ & 0.50 & -0.0013 & 0.0705 & 0.0661 & 0.50 & -0.0120 & 0.0883 & 0.0852 \\ 
  &$d_1$ & 0.50 & -0.0095 & 0.1005 & 0.1020 & 1.00 & -0.0014 & 0.2746 & 0.2505 \\ 
  &$d_2$ & 0.30 & -0.0067 & 0.0606 & 0.0599 & 0.30 & 0.0334 & 0.1427 & 0.1343 \\ 
  &$\gamma_{11}$ & 0.95 & 0.0002 & 0.0154 & 0.0147 & 0.90 & 0.0045 & 0.0513 & 0.0437 \\ 
  &$\gamma_{21}$ & 0.05 & 0.0030 & 0.0167 & 0.0152 & 0.10 & 0.0019 & 0.0379 & 0.0329 \\ 
  &$\gamma_{12}$ & 0.05 & -0.0002 & 0.0154 & 0.0147 & 0.10 & -0.0045 & 0.0513 & 0.0437 \\ 
  &$\gamma_{22}$ & 0.95 & -0.0030 & 0.0167 & 0.0152 & 0.90 & -0.0019 & 0.0379 & 0.0329 \\ 
  &$\delta_{1}$ & 0.50 & 0.0146 & 0.0901 & 0.0886 & 0.50 & 0.0290 & 0.1051 & 0.1029 \\ 
  &$\delta_{2}$ & 0.50 & -0.0146 & 0.0901 & 0.0886 & 0.50 & -0.0290 & 0.1051 & 0.1029 \\
  
  \noalign{\vskip 3mm} 
$1000$ & $a_1$ & -0.50 & 0.0017 & 0.0512 & 0.0503 & 0.20 & 0.0034 & 0.1055 & 0.1036 \\ 
  &$a_2$ & 0.40 & -0.0020 & 0.0478 & 0.0452 & 0.40 & -0.0021 & 0.0672 & 0.0654 \\ 
  &$b_1$ & -0.35 & 0.0033 & 0.0784 & 0.0774 & 0.30 & 0.0116 & 0.0685 & 0.0647 \\ 
  &$b_2$ & 0.50 & 0.0059 & 0.0474 & 0.0459 & 0.50 & -0.0006 & 0.0642 & 0.0634 \\ 
  &$d_1$ & 0.50 & -0.0060 & 0.0712 & 0.0728 & 1.00 & -0.0271 & 0.1820 & 0.1789 \\ 
  &$d_2$ & 0.30 & -0.0123 & 0.0408 & 0.0416 & 0.30 & 0.0063 & 0.0926 & 0.0928 \\ 
  &$\gamma_{11}$ & 0.95 & 0.0010 & 0.0106 & 0.0104 & 0.90 & 0.0085 & 0.0320 & 0.0306 \\ 
  &$\gamma_{21}$ & 0.05 & 0.0015 & 0.0111 & 0.0106 & 0.10 & -0.0010 & 0.0219 & 0.0224 \\ 
  &$\gamma_{12}$ & 0.05 & -0.0010 & 0.0106 & 0.0104 & 0.10 & -0.0085 & 0.0320 & 0.0306 \\ 
  &$\gamma_{22}$ & 0.95 & -0.0015 & 0.0111 & 0.0106 & 0.90 & 0.0010 & 0.0219 & 0.0224 \\ 
  &$\delta_{1}$ & 0.50 & 0.0125 & 0.0658 & 0.0659 & 0.50 & 0.0268 & 0.0775 & 0.0768 \\ 
  &$\delta_{2}$ & 0.50 & -0.0125 & 0.0658 & 0.0659 & 0.50 & -0.0268 & 0.0775 & 0.0768 \\ 
  
  \bottomrule
\end{tabular}
}
\label{table:sim}
\end{table}

 \begin{table}[ht]
\caption{Comparison of models, not including covariates}
\centering
\scalebox{0.8}{
\begin{tabular}{l c c c c c}
\toprule
m & $p$ & df & MSE  & AIC & BIC\\
\midrule
 1 & 3 & 390 & 1.22 & 1469.40 & 1481.32 \\ 
  2 & 8 & 385 & 1.04 & 1450.05 & 1481.85 \\ 
  3 & 13 & 380 & 1.00 & 1460.76 & 1512.42 \\ 
  \bottomrule
\end{tabular}
}
\label{table:comp1}
\end{table}

\begin{table}[ht]
\caption{Parameter estimates. Not including covariates}
\centering
\scalebox{0.8}{
% \begin{tabular}{l l@{\hspace{.5cm}} c c c c@{\hspace{.5cm}} c c c c@{\hspace{.5cm}} c c c c}
\begin{tabular}{l c c }
\toprule
Parameter & Estimate & $\text{St. error}$\\
\midrule
\multicolumn{3}{c}{m = 1}\\
\midrule
$a$ & 0.5691 & 0.0553 \\ 
  $b$ & 0.4075 & 0.0539 \\ 
  $d$ & -0.0688 & 0.0255 \\ 
\midrule
\multicolumn{3}{c}{m = 2}\\
\midrule
$a_1$ & 0.9963 & 0.0093 \\ 
  $a_2$ & 0.5779 & 0.1141 \\ 
  $b_1$ & -0.0386 & 0.0170 \\ 
  $b_2$ & 0.4099 & 0.1069 \\ 
  $d_1$ & 0.0270 & 0.0125 \\ 
  $d_2$ & -0.0252 & 0.0177 \\ 
  $\gamma_{11}$ & 0.9654 & 0.0156 \\ 
  $\gamma_{21}$ & 0.0520 & 0.0289 \\ 
  $\gamma_{12}$ & 0.0346 & 0.0156 \\ 
  $\gamma_{22}$ & 0.9480 & 0.0289 \\ 
  $\delta_{1}$ & 0.6005 & 0.1120 \\ 
  $\delta_{2}$ & 0.3995 & 0.1120 \\ 
\midrule
\multicolumn{3}{c}{m = 3}\\
\midrule
$a_1$ & 0.9479 & 0.0331 \\ 
  $a_2$ & 0.0273 & 0.4309 \\ 
  $a_3$ & -0.1963 & 0.3864 \\ 
  $b_1$ & 0.0404 & 0.0298 \\ 
  $b_2$ & 0.3506 & 0.1587 \\ 
  $b_3$ & 0.5150 & 0.1531 \\ 
  $d_1$ & -0.0191 & 0.0116 \\ 
  $d_2$ & 0.5304 & 0.4326 \\ 
  $d_3$ & 1.2612 & 0.7426 \\ 
  $\gamma_{11}$ & 0.9764 & 0.0232 \\ 
  $\gamma_{21}$ & 0.0270 & 0.0419 \\ 
  $\gamma_{31}$ & 0.0449 & 0.0296 \\ 
  $\gamma_{12}$ & 0.0236 & 0.0232 \\ 
  $\gamma_{22}$ & 0.9478 & 0.0372 \\ 
  $\gamma_{32}$ & 0.0000 & 0.0000 \\ 
  $\gamma_{13}$ & 0.0000 & 0.0000 \\ 
  $\gamma_{23}$ & 0.0253 & 0.0252 \\ 
  $\gamma_{33}$ & 0.9551 & 0.0296 \\ 
  $\delta_{1}$ & 0.5859 & 0.1809 \\ 
  $\delta_{2}$ & 0.2649 & 0.1624 \\ 
  $\delta_{3}$ & 0.1492 & 0.0981 \\ 
  \bottomrule
\end{tabular}
}
\label{table:pars}
\end{table}

 \begin{table}[ht]
\caption{Comparison of models, including covariates one-by-one}
\centering
\scalebox{0.8}{
\begin{tabular}{l c c c c c c c}
\toprule
covariate& MSE & AIC & BIC  & Significant in regime 1 & Significant in regime 2\\
\midrule
indpro & 1.12 & 1437.14 & 1476.88 & YES & NO \\ 
  permit & 0.99 & 1454.14 & 1493.88 & NO & NO \\ 
  ppifgs & 0.98 & 1447.76 & 1487.49 & NO & YES \\ 
  ppieng & 1.00 & 1453.33 & 1493.07 & YES & NO \\ 
  unrate & 0.97 & 1449.97 & 1489.71 & NO & YES \\ 
  baa & 0.91 & 1446.10 & 1485.84 & NO & YES \\ 
  fedfunds & 0.91 & 1451.63 & 1491.37 & NO & NO \\ 
  gs10 & 1.06 & 1454.36 & 1494.10 & NO & NO \\ 
  SP500ret & 0.95 & 1446.92 & 1486.65 & NO & YES \\ 
  SP500vol & 1.02 & 1452.69 & 1492.43 & NO & YES \\ 
  \bottomrule
\end{tabular}
}
\label{table:cov1}
\end{table}

\appendix

%%%%%%%%%%%%%%%%%%%%%%%%%%%%%%%%%%%%%%%%%%%%%%%%%%%%%%%%%%%%%%%%%%%%%%%%%%%%%%%%%%%%%%%%%%
\section{Impementation details}
\label{sec:implementation details}
%%%%%%%%%%%%%%%%%%%%%%%%%%%%%%%%%%%%%%%%%%%%%%%%%%%%%%%%%%%%%%%%%%%%%%%%%%%%%%%%%%%%%%%%%%

The MS-PLLAR EHG algorithm is implemented using the free and open source R (\cite{R}) package Template Model Builder (\textit{TMB}, \cite{TMB}), which is designed for estimating complex nonlinear models. The parameter constraints $\gamma_{ij}\in (0,1)$ and $\sum_{j = 1}^m \gamma_{ij} = 1$ are handled by maximizing a reparametrized version of the quasi log-likelihood $\log L^*(\psi)$, where $\psi = g^{-1}(\theta)$ represent a set of unconstrained parameters. By defining $\log L^* (\psi)$ as a C++ template function TMB provides as R output the likelihood, it's exact gradient and (if needed) it's exact Hessian, where the gradient and Hessian is obtained by automatic differentiation (\cite{Fournier2012}).  The exact gradient allows us to improve the speed and accuracy of the QMLE's by using a gradient-based optimization method, in our case we opted for the R-routine \textit{nlminb}. By reporting $\theta = g(\psi)$ in the C++ template, TMB can provide R-output of model estimates and accompanying standard deviations. The standard deviations are obtained by combining the delta-method and the exact Hessian of $\log L^*(\psi)$ evaluated at the maximum $\hat{\psi}$:

\begin{equation}
\hat{\Sigma}   = - \nabla g(\hat{\psi})\left(\nabla^2\log L^*(\hat{\psi})\right)^{-1}\nabla g(\hat{\psi})^\prime
\label{eq:deltamethod}
\end{equation}
The C++ template function is available from the authors upon request.  

%%%%%%%%%%%%%%%%%%%%%%%%%%%%%%%%%%%%%%%%%%%%%%%%%%%%%%%%%%%%%%%%%%%%%%%%%%%%%%%%%%%%%%%%%%
\subsection{Initialization of the algorithm}
\label{start}
%%%%%%%%%%%%%%%%%%%%%%%%%%%%%%%%%%%%%%%%%%%%%%%%%%%%%%%%%%%%%%%%%%%%%%%%%%%%%%%%%%%%%%%%%%

Implementation of model \eqref{eq:msloglinear} for $m=1$ is investigated in \cite{liboschik2015}, where it is suggested that preferable starting values of $Y_0$ and $\eta_0$ are their respective marginal expectations, assuming a model without covariate effect. For model \eqref{eq:msloglinear} with $m = 1$ and no covariate effects it approximately holds (see \cite{liboschik2015}) that
\begin{equation}
E\left(\log(Y_t + 1)\right) = E(\eta_t) = \frac{d}{1 - a - b}
\end{equation}
Thus, for $m>1$ it is natural to let $\eta_{0\mid S^*_{1} = j, \Omega_0} = E(\eta_t) \approx \sum_{i = 1}^m \delta_i d_i/(1 - a_i - b_i)$, $j = 1, \dots, m^2$ be the elements of $\Lambda_0$ and let $Y_0 =  E(Y_t) \approx \exp(\sum_{i = 1}^m \delta_i d_i/(1 - a_i - b_i))$. The stationary distribution of $S_{t}$ , $\delta = (\delta_1 \dots, \delta_{m})$, is given by $\delta = \mathbf{1}_{m}(I_{m} - \Gamma^* + U_{m})^{-1}$, where $\mathbf{1}_{m}$ is a row vector of ones, $I_{m}$ is the $m\times m$ identity matrix, and $U_{m}$ is the $m\times m$ matrix of ones.

The initialization also requires input of $(P(S^*_{1} = 1\mid\Omega_{0}), \dots, P(S^*_{1} = m^2\mid\Omega_{0})$. We assume $(P(S^*_{1} = 1\mid\Omega_{0}), \dots, P(S^*_{1} = m^2\mid\Omega_{0})) = \delta^{*}$, where $\delta^{*}$ is the stationary distribution  of $S_{t}^*$, and analogously to $S_t$, the stationary distribution of $S_{t}^*$ is given by $\delta^* = \mathbf{1}_{m^2}(I_{m^2} - \Gamma^* + U_{m^2})^{-1}$. 

%%%%%%%%%%%%%%%%%%%%%%%%%%%%%%%%%%%%%%%%%%%%%%%%%%%%%%%%%%%%%%%%%%%%%%
% \subsection{Numerical stability}
%%%%%%%%%%%%%%%%%%%%%%%%%%%%%%%%%%%%%%%%%%%%%%%%%%%%%%%%%%%%%%%%%%%%%%

%%%%%%%%%%%%%%%%%%%%%%%%%%%%%%%%%%%%%%%%%%%%%%%%%%%%%%%%%%%%%%%%%%%%%% 
% \section{Standardadized sampling distribution of $\hat{\theta}$}
%%%%%%%%%%%%%%%%%%%%%%%%%%%%%%%%%%%%%%%%%%%%%%%%%%%%%%%%%%%%%%%%%%%%%%

\end{document}